\documentclass[11pt]{article}

\usepackage[final]{acl}

\usepackage{times}
\usepackage{latexsym}

\usepackage[T1]{fontenc}

\usepackage[utf8]{inputenc}

\usepackage{microtype}

\usepackage{inconsolata}

\usepackage{booktabs}   
\usepackage{multirow}   
\usepackage{graphicx}   
\usepackage{makecell}   
\usepackage{amssymb}    
\usepackage{pifont}     
\usepackage{amsmath}
\usepackage{algorithm}
\usepackage{algpseudocode}
\usepackage{makecell}
\usepackage{placeins}
\usepackage{tabularx}
\usepackage{algorithm}
\usepackage{algpseudocode}
\usepackage[most]{tcolorbox}
\usepackage{marvosym}
\usepackage{amssymb}
\usepackage{svg}

\usepackage{siunitx}
\usepackage{xcolor} 

\newcommand{\greencheck}{{\color[HTML]{228B22}\ding{51}}} 
\newcommand{\redcross}{{\color[HTML]{B22222}\ding{55}}}   

\newcommand\blfootnote[1]{%
  \begingroup
  \renewcommand\thefootnote{}\footnote{#1}%
  \addtocounter{footnote}{-1}%
  \endgroup
}
\newcommand{\best}[1]{\textbf{#1}}

\newcommand{\equalheart}{\textsuperscript{$\heartsuit$}}

\title{AROMA: Augmented Reasoning Over a Multimodal Architecture for Virtual Cell Genetic Perturbation Modeling}

\author{
  \textbf{Zhenyu Wang\textsuperscript{1,2,}\equalheart\textsuperscript{*}},
  \textbf{Geyan Ye\textsuperscript{1,}\equalheart\textsuperscript{\Letter}}, 
  \textbf{Wei Liu\textsuperscript{1}},
  \textbf{Man Tat Alexander Ng\textsuperscript{1}}
\\
\\
  \textsuperscript{1}AI for Life Sciences Lab, Tencent, Shenzhen, China
\\
  \textsuperscript{2}Shenzhen International Graduate School, Tsinghua University, Shenzhen, China
\\
  \href{mailto:blazerye@tencent.com}{blazerye@tencent.com}
  \quad
  \href{mailto:zhenyuwa24@mails.tsinghua.edu.cn}{zhenyuwa24@mails.tsinghua.edu.cn}
}

\begin{document}
\maketitle
\blfootnote{
  \hspace{-1.7em} 
  \begin{tabular}{@{}l}
    {\scalebox{1.0}{$\heartsuit$}} Equal Contribution. \\
    {\scalebox{1.0}{\Letter}} Corresponding Author \& Project Lead. \\
    {\scalebox{1.0}{*}} Work was done when Zhenyu Wang worked as an intern at \\ Tencent.
  \end{tabular}
}
\begin{abstract}

Virtual cell modeling predicts molecular state changes under genetic perturbations in silico, which is essential for biological mechanism studies. However, existing approaches suffer from unconstrained reasoning, uninterpretable predictions, and retrieval signals that are weakly aligned with regulatory topology. To address these limitations, we propose AROMA, an Augmented Reasoning Over a Multimodal Architecture for virtual cell genetic perturbation modeling. AROMA integrates textual evidence, graph-topology information, and protein sequence features to model perturbation-target dependencies, and is trained with a two-stage optimization strategy to yield predictions that are both accurate and interpretable. We also construct two knowledge graphs and a perturbation reasoning dataset, PerturbReason, containing more than 498k samples, as reusable resources for the virtual cell domain. Experiments show that AROMA outperforms existing methods across multiple cell lines, and remains robust under zero-shot evaluation on an unseen cell line, as well as in knowledge-sparse, long-tail scenarios. Overall, AROMA demonstrates that combining knowledge-driven multimodal modeling with evidence retrieval provides a promising pathway toward more reliable and interpretable virtual cell perturbation prediction.  Model weights are available at \url{https://huggingface.co/blazerye/AROMA}. Code is available at \url{https://github.com/blazerye/AROMA}.

\end{abstract}

\section{Introduction}

Cells are the fundamental functional units of living systems, and predicting how cellular molecular states change following genetic interventions is a central problem in modern biology \citep{karr2012whole, bunne2024build}. Virtual cell modeling aims to address this problem in silico by forecasting perturbation responses under a specific cellular context. In this work, we focus on genetic perturbation prediction at the gene level: given a cell line context, a perturbation gene, and a target gene, the model predicts whether the target gene is upregulated, downregulated, or remains non-differentially expressed. This task formulation is motivated by practical needs in biological hypothesis generation and intervention analysis. However, it remains challenging because perturbation effects are highly context-dependent and can propagate through multi-step regulatory cascades.

\begin{figure}[t!]
    \centering
    \includegraphics[width=0.479\textwidth]{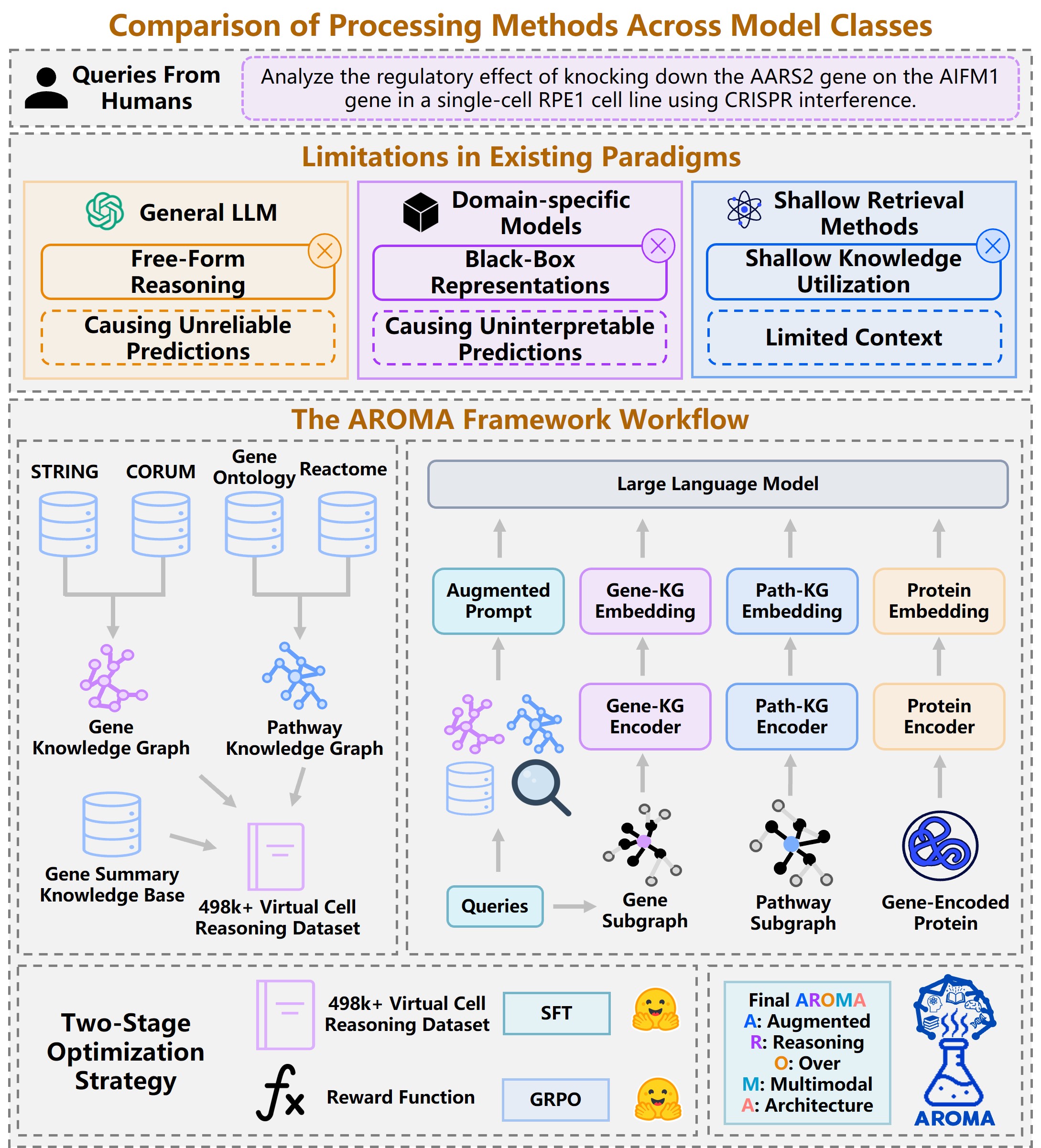}
    \caption{Limitations of existing virtual cell modeling methods and an overview of the AROMA framework.}
    \label{fig:workflow}
\end{figure}

Despite rapid progress in virtual cell modeling, existing approaches still face three recurring limitations in perturbation prediction and reasoning, as summarized in the upper panel of Figure~\ref{fig:workflow}. \textbf{First, unreliable free-form reasoning under interventions.} General Large Language Models \citep{dubey2024llama, guo2025deepseek, hurst2024gpt, comanici2025gemini, yang2025qwen3} can generate fluent explanations, but without explicit biological constraints, they often yield biologically implausible or unreliable predictions for genetic perturbations. Domain-tuned language models \citep{istrate2025rbio1, phillips2025synthpert} trained on synthetic or weakly supervised reasoning traces can improve plausibility, but they may inherit supervision noise and thus remain less reliable. \textbf{Second, prediction without human-interpretable reasoning.} Many existing foundation models and cell-state predictors \citep{roohani2024predicting, chen2024genept, cui2024scgpt, hao2024large, adduri2025predicting} primarily return differential expression scores or labels, without explicit rationales for why a target gene changes under a given perturbation, limiting their value for hypothesis generation and intervention analysis where transparent justifications are required. \textbf{Third, retrieval is not structurally grounded in regulatory mechanisms.} Retrieval-augmented approaches \citep{wu2025contextualizing, wei2025vcworld} access external knowledge, but retrieved evidence is loosely structured, topology-agnostic, and does not model regulatory directionality or multi-step propagation. Consequently, retrieval often offers general functional context but cannot be systematically aligned with mechanistic pathways from a perturbation gene to a downstream target, limiting mechanistically grounded perturbation reasoning.

To address these issues, we propose AROMA, an Augmented Reasoning Over a Multimodal Architecture for virtual cell genetic perturbation modeling. The core idea of AROMA is that perturbation prediction should be grounded in structured, query-specific biological evidence, explicitly modeling the dependency between the perturbation gene and the target gene. The overall framework is illustrated in the lower panel of Figure~\ref{fig:workflow}, which summarizes the data construction, model design, and training pipeline. Concretely, AROMA constructs two complementary knowledge resources that provide topology-grounded biological evidence: a \textbf{Gene Knowledge Graph (Gene-KG)} capturing gene-gene associations, and a \textbf{Pathway Knowledge Graph (Path-KG)} encoding biological interaction structures. Given a perturbation query, the model retrieves functional descriptions and connectivity evidence centered on the perturbation and target genes, constraining the prediction process to biological contexts consistent with regulatory mechanisms. Building on this structured evidence, AROMA further constructs both structural and molecular level representations for the perturbation and target genes, and models their relationship through an interaction mechanism, enabling query-conditioned multimodal reasoning. As a result, predictions are no longer driven solely by text, but are supported by explicit structural and molecular evidence. Finally, we construct a reasoning dataset \textbf{PerturbReason}, and optimize the model using a two-stage training strategy to improve predictive accuracy while providing interpretable explanations in perturbation predictions.

Building on the above, we summarize our main contributions as follows:

\noindent\textbullet~\textbf{Multimodal Knowledge-Driven Framework for Virtual Cell Perturbation Reasoning.} We propose AROMA, which integrates retrieved biological knowledge with molecular-level information to model perturbation effects. Unlike label-only predictors, AROMA also generates human-interpretable reasoning paths explaining why a perturbation affects downstream target genes.

\noindent\textbullet~\textbf{Constructing Two Knowledge Graphs and a Perturbation Reasoning Dataset.}
We construct two scientific knowledge graphs, namely the Gene Knowledge Graph and the Pathway Knowledge Graph, as reusable resources that capture gene functional relationships and regulatory interaction topology. Building on these resources, we curate the PerturbReason perturbation reasoning dataset, contributing structured resources to the virtual cell modeling community.

\noindent\textbullet~\textbf{State-of-the-art performance and robust generalization.} Across multiple cell lines, AROMA consistently outperforms existing models, while achieving strong zero-shot generalization to unseen cell line and remaining robust on knowledge-sparse, long-tail genes.

\section{Related Work}

\subsection{Genetic Perturbation Prediction}

Virtual cell modeling aims to predict cellular responses under genetic interventions. Earlier work relied on mechanistic whole-cell simulations \citep{karr2012whole}, whereas recent methods learn perturbation responses from large-scale perturbation datasets \citep{bunne2024build}, and mainly differ in how they encode biological priors and whether they provide explicit perturbation-to-target evidence. Some methods inject structured priors through regulatory networks or molecular interaction graphs, such as GenePT \citep{chen2024genept} and GEARS \citep{roohani2024predicting}. Others focus on representation learning or generative modeling of post-perturbation expression states across cellular contexts, including scFoundation \citep{hao2024large} and scGPT \citep{cui2024scgpt} and perturbation-specific models such as CPA \citep{lotfollahi2023predicting}, STATE \citep{adduri2025predicting}, and CellFlow \citep{klein2025cellflow}. However, these approaches generally lack human-interpretable mechanistic explanations, limiting their practical utility in biological analysis and hypothesis validation. Recently, retrieval-augmented and LLM-based frameworks, such as SUMMER \citep{wu2025contextualizing} and VCWorld \citep{wei2025vcworld}, incorporate external textual evidence for perturbation reasoning. However, retrieval is typically driven by textual similarity and may overlook structured regulatory topology linking perturbation and target genes. Separately, works such as SynthPert \citep{phillips2025synthpert} and rBio-1 \citep{istrate2025rbio1} train models using synthetic or weakly supervised reasoning traces, where supervision noise may lead to unreliable predictions \citep{ji2023survey}. Complementary to these directions, we retrieve descriptions of genes and their regulatory path evidence, and jointly leverage structural and molecular information within a multimodal architecture, thereby enabling evidence-driven and human-interpretable predictions.

\subsection{Multimodal Learning in Biology}

Multimodal learning has become increasingly important in biology, with models integrating sequences, structures, and text \citep{montesinos2024review, zhang2025atomas, fallahpour2025bioreason, yuan2025annotation, kim2024medexqa}. Two modalities are especially relevant here: gene-scale KG structure, encoded by structure-aware graph encoders \citep{velivckovic2017graph, ma2022cross}, and biomolecular sequences, encoded by pretrained models \citep{lin2023evolutionary, ye2025drugassist, elnaggar2021prottrans}. In perturbation prediction, it remains relatively underexplored to jointly leverage structure-aware KG representations and sequence-derived molecular features for modeling perturbation-target gene interactions, motivating our interaction-centric multimodal design.

\subsection{Knowledge Bases and Benchmarks}

Biological knowledge bases provide structured priors about molecular interactions and functional organization. STRING \citep{mering2003string} and CORUM \citep{giurgiu2019corum} curate gene-gene associations, while GO \citep{ashburner2000gene} and Reactome \citep{croft2010reactome} capture pathway-level processes. Gene-level interaction graphs offer local connectivity, whereas pathway resources provide higher-level organization, and are often used as evidence sources in perturbation modeling. In parallel, benchmarks such as PerturbQA  \citep{wu2025contextualizing} support standardized evaluation of perturbation reasoning and generalization.

\begin{figure*}[t]
    \centering
    \includegraphics[width=1.03\textwidth]{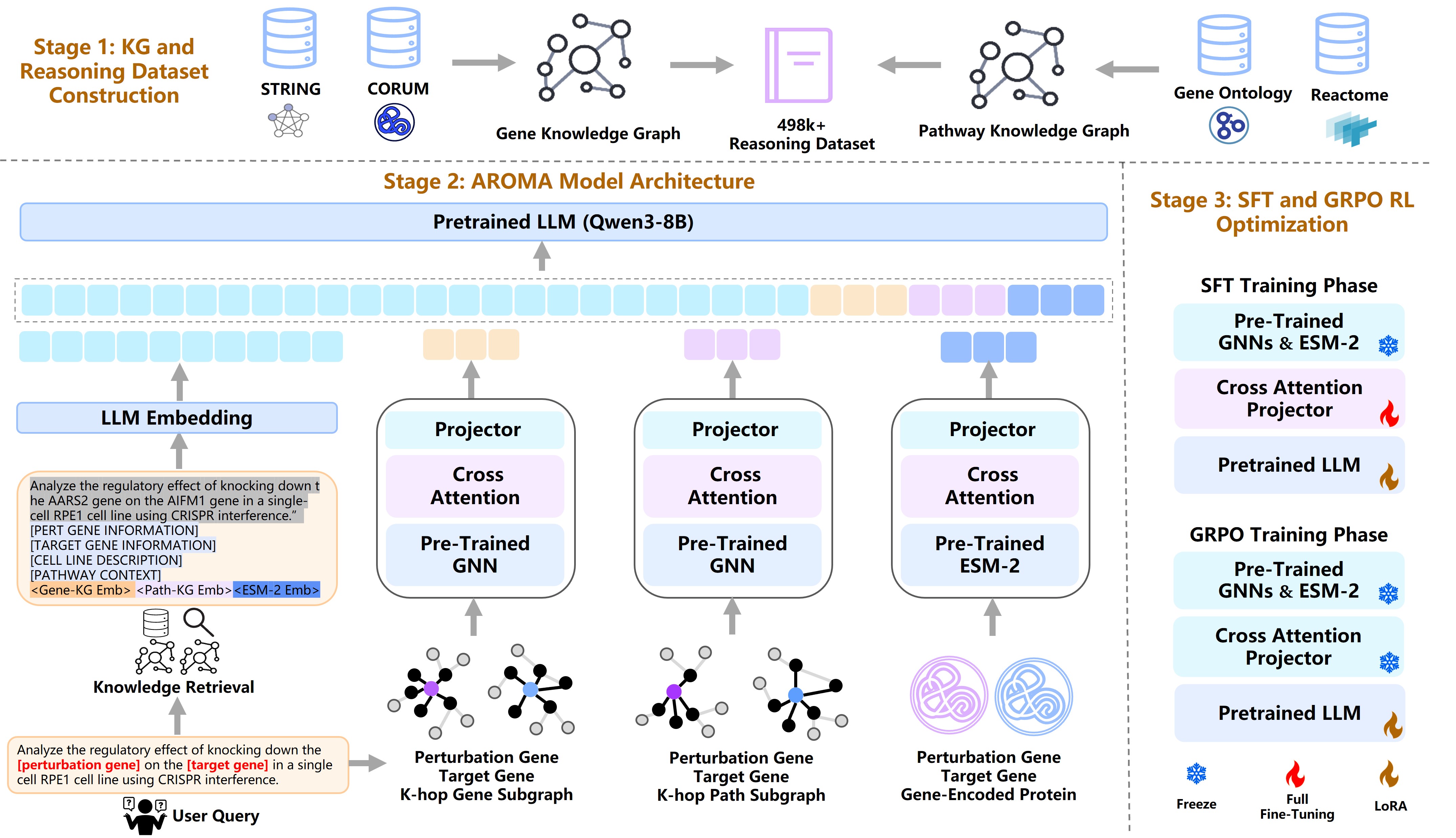}
    \caption{Overview of the AROMA pipeline. An Augmented Reasoning Over a Multimodal Architecture for virtual cell genetic perturbation modeling.}
    \label{fig:aroma_model}
\end{figure*}

\section{Methodology}
\label{sec:method}

\subsection{Problem Definition and Framework Overview}
\label{subsec:problem_def}

This work studies the problem of predicting target gene expression states under genetic perturbations. Given a cell line context $c \in \mathcal{C}$, a perturbation gene $g_p \in \mathcal{G}$, and a target gene $g_t \in \mathcal{G}$, the goal is to predict a differential expression label $y \in \{\text{Up}, \text{Down}, \text{Non-diff}\}$ for each triplet $x = (g_p, g_t, c)$. Model parameters are learned by maximizing the conditional log-likelihood over a training set $\mathcal{D}$:

\begin{equation}
\small
\Theta^* = \arg\max_{\Theta} \sum_{(x,y)\in \mathcal{D}} \log P_{\Theta}(y \mid x)
\end{equation}
where $\mathcal{C}$ and $\mathcal{G}$ denote the sets of cell line contexts and genes, $\Theta$ and $\Theta^{*}$ denotes the model parameters, and $P_{\Theta}(y\mid x)$ denotes the conditional label distribution. Unlike standard classification tasks, this problem requires the model not only to output predictions but also to produce biologically meaningful explanations. To address this problem, we propose AROMA, whose overall pipeline is illustrated in Figure~\ref{fig:aroma_model} and divided into three stages:

\noindent\textbullet~\textbf{Data stage.} AROMA constructs two complementary knowledge graphs and a large-scale virtual cell reasoning dataset for evidence grounding, as described in Section~\ref{sec:data}.

\noindent\textbullet~\textbf{Modeling stage.} AROMA adopts a retrieval-augmented strategy to incorporate query-relevant information, thereby providing explicit evidence cues for prediction, as described in Section~\ref{sec:retrieval}. In addition, it jointly leverages topological representations learned from graph neural networks (GNN) and protein sequence representations encoded by ESM-2, and applies a cross-attention module to explicitly model perturbation-target gene dependencies across modalities, as described in Section~\ref{sec:model}. 

\noindent\textbullet~\textbf{Training stage.} AROMA first performs multimodal supervised fine-tuning (SFT), and is then further optimized with Group Relative Policy Optimization (GRPO) reinforcement learning to enhance predictive performance while generating biologically meaningful explanations, as described in Section~\ref{sec:two_stage}.

\begin{figure*}[t]
    \centering
    \includegraphics[width=1.03\textwidth]{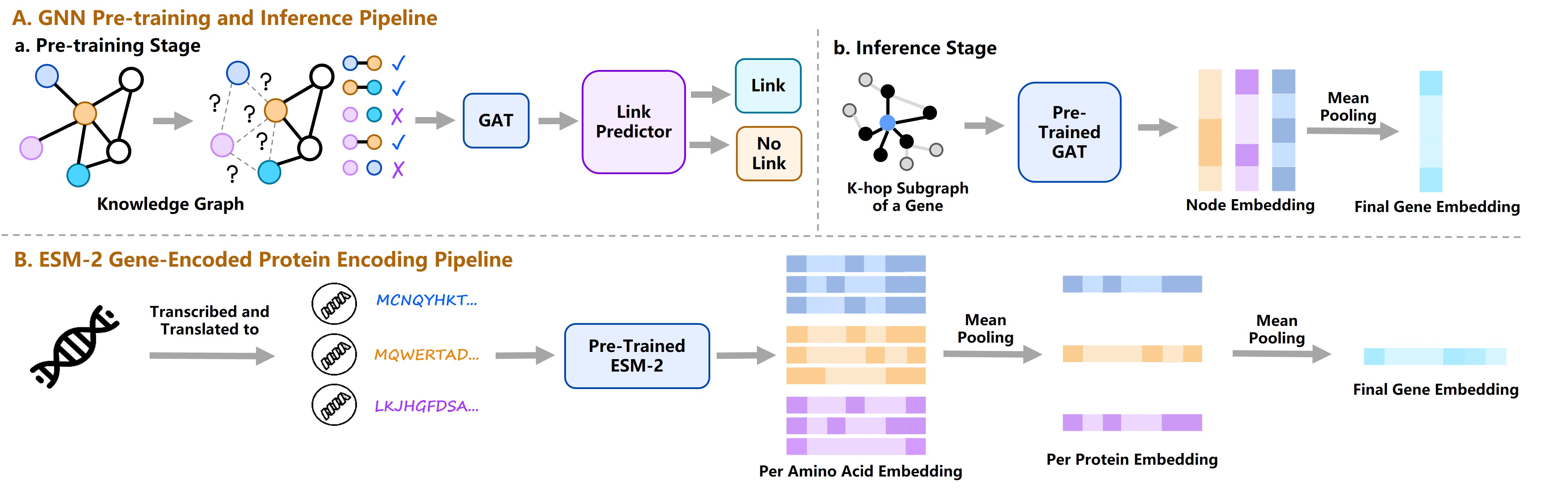}
    \caption{\textbf{Structural and sequence encoders in AROMA.}
    \textbf{A:} A pretrained GAT on the Gene-KG and the Path-KG encodes gene-centered subgraphs to obtain structural representations.
    \textbf{B:} A pretrained ESM-2 model encodes the amino-acid sequence of each gene's protein to obtain sequence representations.}
    \label{fig:results}
\end{figure*}

\subsection{Construction of Knowledge Graphs and Reasoning Dataset}
\label{sec:data}

To provide structured biological evidence for gene perturbation prediction, we construct two knowledge graphs and a perturbation reasoning dataset: the \textbf{Gene Knowledge Graph (Gene-KG)}, the \textbf{Pathway Knowledge Graph (Path-KG)}, and \textbf{PerturbReason}. Gene-KG integrates STRING \citep{mering2003string} and CORUM \citep{giurgiu2019corum} into a gene-level association network with over \textbf{18k nodes and 700k edges}, while Path-KG is constructed from Gene Ontology \citep{ashburner2000gene} and Reactome \citep{croft2010reactome}, whose nodes span genes, functional terms, and biological process entities, with over \textbf{80k nodes and 400k edges}. The two graphs are complementary in granularity and coverage, jointly providing multi-level evidence, where edges are treated as associative rather than explicit causal relations. Building on these resources, we construct PerturbReason, a reasoning dataset with more than 498k samples, based on the PerturbQA dataset \citep{wu2025contextualizing}. For each instance, we retrieve gene functional descriptions, regulatory paths from the knowledge graphs, and cell line information, and organize the evidence into an instruction-style input. Then, based on the above evidence, we use DeepSeek-R1 \citep{guo2025deepseek} to generate label-consistent reasoning traces and corresponding predictions, forming reasoning data for model training. To reduce supervision noise, we retain only instances with label-consistent answers and structurally valid reasoning. Details of the data construction prompt templates, data construction procedure, and data quality analysis are provided in the Appendix~\ref{app:data} and Appendix~\ref{app:reasoning_quality}. Together, these resources provide reusable knowledge and training data for virtual cell domain.

\subsection{Retrieval-Augmented Contextualization}
\label{sec:retrieval}

To provide the model with richer context, we construct a retrieval-augmented textual input \(X_{\text{text}}\) by combining the query instance \(q\) with three types of external evidence: gene functional descriptions \(T_{\text{desc}}\), regulatory paths between genes \(T_{\text{path}}\), and cell line descriptions \(T_{\text{cell}}\). The components are mapped into a unified text representation by \(\Phi(\cdot)\):

\begin{equation}
\small
    X_{\text{text}} = \Phi\bigl(q, T_{\text{desc}}, T_{\text{path}}, T_{\text{cell}}\bigr)
    \label{eq:xtext}
\end{equation}

\noindent\textbullet~\textbf{Gene functional descriptions} \(\boldsymbol{T}_{\textbf{\text{desc}}}\). In the SUMMER knowledge base \citep{wu2025contextualizing}, each gene has two predefined functional descriptions: characterizing its roles as a perturbation gene and as a target gene. We automatically retrieve these node-level functional summaries to obtain \(T_{\text{desc}}\). 

\noindent\textbullet~\textbf{Regulatory paths between genes} \(\boldsymbol{T}_{\textbf{\text{path}}}\). To expose explicit regulatory structure, we use breadth-first search \citep{bundy1984breadth} on the Path-KG to extract up to three shortest paths, measured by hop count, from the perturbation gene \(g_p\) to the target gene \(g_t\), denoted as \(T_{\text{path}}\), if no valid path exists, \(T_{\text{path}}\) is set to be empty. 

\noindent\textbullet~\textbf{Cell line descriptions} \(\boldsymbol{T}_{\textbf{\text{cell}}}\). We construct a cell line description repository based on Wikipedia entries and retrieve the corresponding cell line descriptions \(T_{\text{cell}}\), enabling the model to perform reasoning under the appropriate cellular context.

\subsection{Multimodal Interaction Encoding}
\label{sec:model}

While retrieval-based textual context provides useful semantic cues for perturbation prediction, structural and molecular representations further enrich modeling of perturbation-target gene relationships. To this end, we introduce a multimodal interaction encoding module that combines graph structure and protein sequences to form perturbation-target gene interaction features.

\noindent\textbf{Structure-Aware Graph Encoders.} For the structural modality, we train two architecturally identical but parameter-independent GAT encoders on the Gene-KG and the Path-KG, respectively. With node states $h_i^{(l)}$ at layer $l$, GAT updates representations via neighborhood attention aggregation:

\begin{equation}
\small
h_i^{(l+1)} = \text{ELU}\Big( \sum_{j \in \mathcal{N}(i)} \alpha_{ij} W^{(l)} h_j^{(l)} \Big)
\end{equation}
where $\mathcal{N}(i)$ denotes the neighborhood of node $i$, $\alpha_{ij}$ is the attention weight, $W^{(l)}$ is a layer-wise projection, and $\mathrm{ELU}(\cdot)$ is the activation function. We first pre-train the two GAT encoders on their respective graphs using an edge-prediction objective, as illustrated in Figure~\ref{fig:results}A(a). Specifically, given node embeddings $z_u$ and $z_v$, we first compute an edge score $s_{uv}$ that predicts whether the candidate edge $(u,v)$ exists:
\begin{equation}
\small
s_{uv} = \mathrm{MLP}\bigl([z_u; z_v]\bigr),
\end{equation}
where $[z_u;z_v]$ denotes vector concatenation. The encoder is then trained with a binary cross-entropy loss:
\begin{equation}
\small
\mathcal{L}_{\text{pre}} = -\sum_{(u,v)\in\mathcal{E}} \log \sigma(s_{uv}) 
-\sum_{(u,v)\notin\mathcal{E}} \log\bigl(1-\sigma(s_{uv})\bigr)
\end{equation}
where $\mathcal{E}$ is the edge set, $\sigma(\cdot)$ is the sigmoid function. Detailed implementation settings and experimental results of GAT pre-training are provided in the Appendix~\ref{app:gnnsample}, and Appendix~\ref{app:gnn_pretrain}. Then, at inference stage, as shown in Figure~\ref{fig:results}A(b), for each query $(g_p,g_t,c)$, we extract $k$-hop subgraphs around $g_p$ and $g_t$ from both knowledge graphs. We encode subgraph nodes with the corresponding pre-trained GAT and mean-pooling the node representations to obtain perturbation and target-specific embeddings in each modality, denoted as $H_{\text{gene}}^{(p)}, H_{\text{gene}}^{(t)}$ and $H_{\text{path}}^{(p)}, H_{\text{path}}^{(t)}$. The choice of $k$ is discussed in the Appendix~\ref{app:subgraph_range}.

\noindent\textbf{Protein Sequence Encoder.} For the sequence modality, as shown in Figure~\ref{fig:results}B, we represent each gene by its encoded protein sequence. Using frozen ESM-2, we encode protein sequences $S_p$ and $S_t$ and mean-pooling token states to obtain  gene-level embeddings:

\begin{equation}
\small
h_p^{\text{prot}}=\mathrm{MP}\!\left(\Phi_{\text{esm}}(S_p)\right)\quad
h_t^{\text{prot}}=\mathrm{MP}\!\left(\Phi_{\text{esm}}(S_t)\right)
\end{equation}
where $\Phi_{\text{esm}}(\cdot)$ is frozen ESM-2 and $\mathrm{MP}(\cdot)$ denotes mean-pooling over tokens. We keep ESM-2 frozen to directly reuse its general-purpose protein representations from large-scale pretraining.

\noindent\textbf{Cross-Modal Interaction Features.} To explicitly model perturbation-target gene interactions, we apply modality-specific cross-attention for  $m \in \{\text{gene}, \text{path}, \text{protein}\}$, using perturbation as query and target as key and value. Interaction features are computed by scaled dot-product attention:
\begin{equation}
\small
z_{\text{inter}}^m =
\text{Softmax}\!\left(\frac{Q_{m,p} K_{m,t}^\top}{\sqrt{d_k}}\right) V_{m,t}
\end{equation}
where $Q_{m,p}$, $K_{m,t}$, and $V_{m,t}$ are linear projections and $d_k$ is the key dimension. We then project each interaction feature into the language model's token space via a lightweight projector. Let $\Phi^{m}_{\text{proj}}$ be the projector and $e^{m}_{\text{inter}}$ be the embedding of the modality-specific placeholder token. The replacement operation is formalized as
\begin{equation}
\small
e^{m}_{\text{inter}} \leftarrow \Phi^{m}_{\text{proj}}\!\left(z_{\text{inter}}^m\right)\quad m\in\{\text{gene},\text{path},\text{protein}\}
\end{equation}
By injecting these interaction embeddings at predefined positions, the language model receives explicit multimodal perturbation-target gene interaction signals in its input, supporting subsequent reasoning and prediction.

\subsection{Two-Stage Optimization Strategy}
\label{sec:two_stage}

To improve predictive performance and regularize the model's reasoning behavior, we adopt a two-stage training scheme.

\noindent\textbf{Multimodal Supervised Fine-Tuning.} The first stage performs multimodal SFT on a domain-specific perturbation reasoning dataset constructed in this work, denoted as $\mathcal{D}_{\text{sft}}$, enabling the model to align information across modalities and adapt to domain-specific knowledge. Each instance contains the input $X$, a reasoning sequence $Y_{\text{reason}}$, and a final answer $Y_{\text{answer}}$, concatenated into the output $Y$. We optimize the standard autoregressive language-modeling objective:
\begin{equation}
\small
\mathcal{L}_{\text{SFT}}(\Theta)
=
-
\sum_{(X,Y)\in \mathcal{D}_{\text{sft}}}
\sum_{t=1}^{|Y|}
\log P(y_t \mid y_{<t}, X; \Theta)
\end{equation}
where $\Theta$ denotes the model parameters, $y_t$ is the $t$-th token, and $y_{<t}$ is its preceding context. During this stage, the GNN encoders and ESM-2 encoder are frozen, the interaction and projection modules are fully fine-tuned, and the language model is updated via LoRA.

\noindent\textbf{GRPO-Based Reinforcement Learning.} In the second stage, we further refine the model's reasoning with GRPO reinforcement learning, encouraging accurate and well-explained predictions. In this stage, the large language model is fine-tuned with LoRA, while all other modules are frozen. For each instance, we sample multiple reasoning trajectories and define a reward that reflects task accuracy and reasoning quality. Concretely, a reward of $5.0$ is given for correct predictions and $-1.0$ otherwise. An additional $0.5$ is added if the reasoning trace follows the predefined format, and another $0.5$ if the answer contains exactly one valid category. Rewards are normalized within each trajectory group to compute advantages, and the policy is optimized using GRPO to improve reasoning quality and answer consistency. Additional implementation details of the two-stage training procedure are provided in the Appendix~\ref{app:case_study2}.

\begin{table*}[t!]
\centering
\caption{Performance comparison on the genetic perturbation prediction. The reported values are the F1-scores for Non-differential (ND), Up-regulated (Up), and Down-regulated (Down) predictions across four cell lines. }
\label{tab:performance_merged}
\scriptsize
\setlength{\tabcolsep}{0pt}
\begin{tabular*}{\textwidth}{@{\extracolsep{\fill}}llccccccccccccc}
\toprule
\multirow{2}{*}{\textbf{Category}} & \multirow{2}{*}{\textbf{Model}} & \multicolumn{3}{c}{\textbf{K562}} & \multicolumn{3}{c}{\textbf{HepG2}} & \multicolumn{3}{c}{\textbf{Jurkat}} & \multicolumn{3}{c}{\textbf{RPE1}} & \multirow{2}{*}{\textbf{Average}} \\
\cmidrule{3-5} \cmidrule{6-8} \cmidrule{9-11} \cmidrule{12-14}
 & & \textbf{ND} & \textbf{Up} & \textbf{Down} & \textbf{ND} & \textbf{Up} & \textbf{Down} & \textbf{ND} & \textbf{Up} & \textbf{Down} & \textbf{ND} & \textbf{Up} & \textbf{Down} & \\
\midrule
\multirow{5}{*}{General LLMs} 
 & DeepSeek-R1 \citep{guo2025deepseek} & 0.69 & 0.21 & 0.07 & 0.61 & 0.20 & 0.22 & 0.61 & 0.19 & 0.19 & 0.63 & 0.16 & 0.14 & 0.33 \\
 & OpenAI o4-mini \citep{openai2025o3o4systemcard}  & 0.71 & 0.23 & 0.07 & 0.64 & 0.22 & 0.23 & 0.63 & 0.20 & 0.27 & 0.68 & 0.13 & 0.21 & 0.35 \\
 & GPT-5 \citep{openai2025gpt5systemcard} & 0.76 & 0.31 & 0.12 & 0.71 & 0.29 & 0.35 & 0.69 & 0.27 & 0.25 & 0.72 & 0.23 & 0.22 & 0.41 \\
 & Gemini-2.5-pro \citep{googlecloud2025gemini25pro} & 0.74 & 0.29 & 0.13 & 0.69 & 0.27 & 0.26 & 0.68 & 0.25 & 0.22 & 0.69 & 0.21 & 0.21 & 0.39 \\
 & Qwen3-235B \citep{yang2025qwen3} & 0.70 & 0.19 & 0.07 & 0.63 & 0.21 & 0.22 & 0.62 & 0.21 & 0.19 & 0.61 & 0.19 & 0.13 & 0.33 \\
\midrule
\multirow{8}{*}{\shortstack[l]{Domain-Specific\\Foundation Models}} 
 & SynthPert \citep{phillips2025synthpert} & 0.92 & 0.27 & 0.17 & 0.91 & 0.21 & 0.25 & 0.93 & 0.15 & 0.31 & 0.95 & 0.23 & 0.53 & 0.49 \\
 & GAT \citep{velivckovic2017graph} & 0.94 & \underline{0.62} & 0.21 & 0.93 & \underline{0.51} & 0.57 & 0.93 & 0.37 & 0.59 & 0.92 & 0.45 & 0.58 & \underline{0.64} \\
 & STATE \citep{adduri2025predicting} & \underline{0.95} & 0.25 & 0.23 & \underline{0.96} & 0.29 & 0.35 & 0.95 & 0.29 & 0.42 & \underline{0.96} & 0.33 & 0.52 & 0.54 \\
 & GEARS \citep{roohani2024predicting} & 0.94 & 0.37 & \underline{0.26} & 0.95 & 0.26 & 0.42 & \underline{0.96} & 0.17 & 0.57 & 0.92 & 0.37 & 0.21 & 0.53 \\
 & scGPT \citep{cui2024scgpt} & 0.92 & 0.52 & 0.15 & 0.93 & 0.46 & 0.54 & 0.93 & 0.34 & 0.54 & 0.92 & 0.42 & 0.57 & 0.60 \\
 & GenePT-Gene \citep{chen2024genept} & 0.93 & 0.34 & 0.24 & 0.93 & 0.30 & 0.55 & 0.94 & 0.24 & 0.60 & 0.93 & 0.34 & 0.57 & 0.58 \\
 & GenePT-Prot \citep{chen2024genept} & 0.94 & 0.47 & 0.17 & 0.94 & 0.37 & 0.56 & 0.95 & 0.31 & 0.61 & 0.93 & 0.39 & 0.61 & 0.60 \\
 & SUMMER \citep{wu2025contextualizing} & 0.94 & 0.56 & 0.25 & 0.94 & 0.47 & \underline{0.60} & 0.94 & \underline{0.39} & \underline{0.62} & 0.92 & \underline{0.46} & \underline{0.63} & \underline{0.64} \\
\midrule
\textbf{Ours} & \textbf{AROMA} & \textbf{0.96} & \textbf{0.66} & \textbf{0.36} & \textbf{0.97} & \textbf{0.65} & \textbf{0.65} & \textbf{0.97} & \textbf{0.58} & \textbf{0.69} & \textbf{0.97} & \textbf{0.60} & \textbf{0.73} & \textbf{0.73} \\
\bottomrule
\end{tabular*}
\end{table*}

\section{Experiments}

\subsection{Experimental Setup and Research Questions}

To systematically evaluate AROMA on gene perturbation prediction, we formulate five research questions, as summarized below:

\noindent\textbullet~\textbf{Overall Predictive Effectiveness.} Can AROMA outperform existing methods on gene perturbation prediction? Described in Section~\ref{subsec:genetic_perturbation}.

\noindent\textbullet~\textbf{Zero-Shot Generalization.} In a zero-shot setting on an unseen cell line, can AROMA avoid significant performance degradation? Described in Section~\ref{subsec:zero-shot}.

\noindent\textbullet~\textbf{Component Contributions.} Do the two-stage optimization and the retrieval, GNN encoders, and ESM-2 encoder modules lead to performance improvements? Described in Section~\ref{subsec:ablation study}.

\noindent\textbullet~\textbf{Knowledge-Sparse Regions.} On low-popularity and weakly connected genes, can RAG-based textual evidence and multimodal structural features mitigate performance degradation in long-tail scenarios? Described in Section~\ref{subsec:knowledge sparsity}.

\noindent\textbullet~\textbf{GRPO Sampling Sensitivity.} How sensitive is AROMA to the number of sampled reasoning trajectories in GRPO, and what trajectory count leads to the most stable performance? Described in Section~\ref{subsec:grpo_sensitivity}.

In addition, we present a case study with sentence-level source-tracing and biological validity analyses of the model-generated reasoning, detailed in Appendix~\ref{sec:case_study}.

All experiments are conducted on four human cell lines: K562 \citep{andersson1979k562}, HepG2 \citep{arzumanian2021curious}, Jurkat \citep{gioia2018genome}, and RPE1 \citep{spalluto2013evidence}. Because the task is a three-class classification problem with imbalanced labels, we adopt Macro-F1 as the evaluation metric. Details of the data processing procedures and baseline configurations are provided in Appendix~\ref{app:case_study2}.

\subsection{Overall Predictive Performance}
\label{subsec:genetic_perturbation}

In this task, each sample contains a cell-line context, a perturbation gene, and a target gene, and the model predicts whether the target gene is Non-differential (ND), Up-regulated (Up), or Down-regulated (Down) under the perturbation. All methods are trained on the PerturbQA training split and evaluated on its official test split, and the results on the four cell lines are reported in Table~\ref{tab:performance_merged}. AROMA achieves the highest average F1-score and consistently outperforms all comparison methods. General-purpose large language models perform worst due to limited domain-specific inductive bias, while domain-specific models benefit from structured priors and obtain stronger results, but their predictions lack interpretability. In contrast, AROMA integrates multimodal representations with retrieval-augmented evidence and a two-stage training strategy, leading to substantially better overall performance.

\subsection{Zero-shot Generalization}
\label{subsec:zero-shot}

To assess the zero-shot generalization capability of AROMA, we design an evaluation setting: unseen cell line. In this setting, the model is trained on K562, HepG2, and Jurkat and evaluated on RPE1, without any RPE1 training data. For comparison, we also consider a non-zero-shot setting where the model is trained on all four cell lines. The specific experimental results are reported in Table \ref{tab:zero_shot}. The experiments show that, even in this zero-shot scenario, AROMA's overall performance degrades only slightly, and in most cases it still outperforms other competing models (see Table \ref{tab:performance_merged}), indicating strong cross-distribution generalization and robustness of AROMA. Such behavior is desirable in biological discovery, where new perturbations and cellular environments are constantly emerging.

\begin{table}[h]
\centering
\small
\caption{Zero-shot performance comparison on the genetic perturbation prediction in the RPE1.}
\label{tab:zero_shot}
\setlength{\tabcolsep}{4pt} 
\begin{tabular}{lcccc}
\toprule
\textbf{Setting} & \textbf{ND} & \textbf{Up} & \textbf{Down} & \textbf{Avg} \\
\midrule
AROMA (Unseen Cell Line) & 0.96 & 0.57 & 0.66 & 0.73 \\
AROMA (Training) & 0.97 & 0.60 & 0.73 & 0.77 \\
\bottomrule
\end{tabular}
\end{table}

\begin{table*}[t!]
\centering
\caption{Ablation study of different modules. \greencheck \ indicates the module is enabled, 
while \redcross \ indicates it is disabled.}
\label{tab:ablation}
\scriptsize
\setlength{\tabcolsep}{3.6pt}
\renewcommand{\arraystretch}{1.12}

\resizebox{\textwidth}{!}{%
\begin{tabular}{ccccc|ccccccccccccc}
\toprule
\multirow{2}{*}{\makecell{\textbf{Training}\\\textbf{Method}}} &
\multirow{2}{*}{\makecell{\textbf{Reasoning}\\\textbf{Data}}} &
\multirow{2}{*}{\makecell{\textbf{RAG}\\\textbf{Module}}} &
\multirow{2}{*}{\makecell{\textbf{GNN}\\\textbf{Encoders}}} &
\multirow{2}{*}{\makecell{\textbf{ESM-2}\\\textbf{Encoder}}} &
\multicolumn{3}{c}{\textbf{K562}} &
\multicolumn{3}{c}{\textbf{HepG2}} &
\multicolumn{3}{c}{\textbf{Jurkat}} &
\multicolumn{3}{c}{\textbf{RPE1 (Zero-shot)}} &
\multirow{2}{*}{\textbf{Avg}} \\
\cmidrule(lr){6-8} \cmidrule(lr){9-11} \cmidrule(lr){12-14} \cmidrule(lr){15-17}
 & & & & & \textbf{ND} & \textbf{Up} & \textbf{Down} & \textbf{ND} & \textbf{Up} & \textbf{Down} & \textbf{ND} & \textbf{Up} & \textbf{Down} & \textbf{ND} & \textbf{Up} & \textbf{Down} & \\
\midrule

Qwen3-8B & \redcross & \redcross & \redcross & \redcross &
0.52 & 0.17 & 0.06 & 0.52 & 0.12 & 0.16 & 0.54 & 0.07 & 0.20 & 0.57 & 0.09 & 0.15 & 0.26 \\

SFT & \greencheck & \redcross & \redcross & \redcross &
0.94 & 0.59 & 0.31 & 0.95 & 0.53 & 0.55 & 0.96 & 0.43 & 0.57 & 0.95 & 0.45 & 0.54 & 0.65 \\

SFT + GRPO & \greencheck & \redcross & \redcross & \redcross &
0.96 & 0.63 & 0.33 & 0.96 & 0.59 & 0.57 & 0.96 & 0.49 & 0.62 & 0.95 & 0.52 & 0.61 & 0.68 \\

SFT + GRPO & \greencheck & \greencheck & \redcross & \redcross &
0.96 & 0.65 & 0.35 & 0.96 & 0.64 & 0.65 & \best{0.97} & 0.54 & 0.68 & \best{0.96} & 0.55 & 0.63 & 0.71 \\

SFT + GRPO & \greencheck & \redcross & \greencheck & \redcross &
0.96 & 0.65 & 0.35 & 0.96 & 0.62 & 0.62 & \best{0.97} & 0.55 & 0.64 & 0.95 & 0.53 & 0.65 & 0.70 \\

SFT + GRPO & \greencheck & \redcross & \redcross & \greencheck &
0.95 & 0.63 & 0.33 & 0.96 & 0.60 & 0.60 & 0.96 & 0.52 & 0.63 & 0.95 & 0.52 & 0.63 & 0.69 \\

SFT + GRPO & \greencheck & \redcross & \greencheck & \greencheck &
0.96 & 0.65 & 0.36 & 0.96 & 0.62 & 0.63 & \best{0.97} & 0.56 & 0.66 & \best{0.96} & 0.53 & 0.65 & 0.71 \\

SFT + GRPO & \greencheck & \greencheck & \greencheck & \greencheck &
\best{0.97} & \best{0.67} & \best{0.39} & \best{0.97} & \best{0.66} & \best{0.69} &
\best{0.97} & \best{0.57} & \best{0.69} & \best{0.96} & \best{0.57} & \best{0.66} & \best{0.73} \\

\bottomrule
\end{tabular}%
}
\end{table*}

\subsection{Ablation Study}
\label{subsec:ablation study}

We conduct an ablation study to quantify the contribution of each core component of AROMA. For each variant, models are trained and evaluated on the standard K562, HepG2, and Jurkat splits, and are additionally assessed in a zero-shot setting on the held-out RPE1 cell line, where no RPE1 data are used during training. The ablation results are summarized in Table~\ref{tab:ablation}. The vanilla Qwen3-8B backbone yields relatively low Macro-F1, indicating insufficient domain-specific inductive bias for reliable perturbation reasoning. In contrast, SFT on task-specific perturbation reasoning data yields substantial gains, suggesting that injecting biologically grounded knowledge is critical for this task. Building on SFT, GRPO-based reinforcement learning further improves performance, suggesting that using task-level feedback to guide the reasoning process enhances prediction accuracy. Incorporating the retrieval-augmented module yields additional improvements, as query-specific external evidence provides complementary external evidence beyond model-internal knowledge. On the multimodal side, the pretrained GNN encoders provide topology-aware structural representations, while the ESM-2 encoder supplies complementary molecular-level information; removing either leads to a clear performance drop, highlighting their complementary roles. With all components enabled, the full AROMA model achieves the best overall performance, demonstrating the cumulative benefits of integrating structural, molecular, and retrieval-based evidence.

\subsection{Robustness Under Knowledge Sparsity}
\label{subsec:knowledge sparsity}

To evaluate the robustness of AROMA under varying knowledge availability, we perform a stratified analysis by grouping genes according to their popularity and their connectivity. Gene popularity is defined as the frequency of gene mentions in ~26M PubMed~\citep{white2020pubmed} titles and abstracts, and connectivity as node degree in the Path-KG. For each criterion, test genes are ranked and partitioned into High, Mid, and Low groups, corresponding to the top 20\%, middle 40\%, and bottom 40\% of samples. This stratified evaluation uses models trained on the combined training sets of all four cell lines, with RPE1 results summarized in Table~\ref{tab:table4_rpe1} and results for the other cell lines reported in Appendix~\ref{app:robustness_stratified}. The stratified analysis shows that AROMA is more robust in knowledge-sparse regions: from high-popularity or highly connected genes to low-popularity or low-degree genes, its performance degrades much less than the variant without retrieval and multimodal modules. This indicates that AROMA's gains do not mainly arise from memorizing high-frequency or highly connected genes, but from jointly modeling retrieval-augmented evidence and multimodal representations, which helps maintain competitive performance on low-exposure, long-tail genes.

\begin{table}[H]
\centering
\caption{Stratified robustness evaluation on the RPE1.}
\label{tab:table4_rpe1}
\footnotesize
\setlength{\tabcolsep}{1.3pt}
\renewcommand{\arraystretch}{1.15}

\begin{tabular}{lcccccc}
\toprule
\multirow{2}{*}{\textbf{Model}} &
\multicolumn{3}{c}{\textbf{Gene Popularity}} &
\multicolumn{3}{c}{\textbf{Node Degree}} \\
\cmidrule(l{1.2pt}r{1.2pt}){2-4}\cmidrule(l{1.2pt}r{1.2pt}){5-7}
& High & Mid & Low & High & Mid & Low \\
\midrule
\textbf{w/o RAG \& Multimodal}
& 0.69 & 0.62 & 0.56 & 0.72 & 0.63 & 0.57 \\
\textbf{AROMA}
& 0.77 & 0.74 & 0.71 & 0.79 & 0.75 & 0.69 \\
\bottomrule
\end{tabular}
\end{table}

\begin{table*}[t!]
\centering
\caption{Sensitivity analysis of the number of sampled GRPO trajectories.}
\label{tab:grpo_sensitivity}
\small 
\setlength{\tabcolsep}{3.5pt} 
\renewcommand{\arraystretch}{1.11} 

\resizebox{\textwidth}{!}{%
\begin{tabular}{cccccccccccccc}
\toprule
\multirow{2}{*}{\textbf{Model}} & \multicolumn{3}{c}{\textbf{K562}} & \multicolumn{3}{c}{\textbf{HepG2}} & \multicolumn{3}{c}{\textbf{Jurkat}} & \multicolumn{3}{c}{\textbf{RPE1}} & \multirow{2}{*}{\textbf{Average}} \\
\cmidrule(lr){2-4} \cmidrule(lr){5-7} \cmidrule(lr){8-10} \cmidrule(lr){11-13}
& \textbf{ND} & \textbf{Up} & \textbf{Down} & \textbf{ND} & \textbf{Up} & \textbf{Down} & \textbf{ND} & \textbf{Up} & \textbf{Down} & \textbf{ND} & \textbf{Up} & \textbf{Down} & \\
\midrule
AROMA (4 trajectories)  & 0.95 & 0.61 & 0.32 & 0.96 & 0.61 & 0.62 & 0.96 & 0.53 & 0.67 & 0.94 & 0.57 & 0.71 & 0.70 \\
AROMA (8 trajectories)  & 0.95 & 0.64 & 0.35 & 0.96 & 0.64 & \textbf{0.65} & \textbf{0.97} & 0.57 & \textbf{0.69} & \textbf{0.97} & 0.59 & \textbf{0.73} & 0.72 \\
AROMA (16 trajectories) & \textbf{0.96} & \textbf{0.66} & \textbf{0.36} & \textbf{0.97} & \textbf{0.65} & \textbf{0.65} & \textbf{0.97} & \textbf{0.58} & \textbf{0.69} & \textbf{0.97} & \textbf{0.60} & \textbf{0.73} & \textbf{0.73} \\
\bottomrule
\end{tabular}%
}
\end{table*}

\subsection{Sensitivity to the Number of GRPO Trajectories}
\label{subsec:grpo_sensitivity}

To further analyze the sensitivity of AROMA to the number of sampled reasoning trajectories in GRPO, we conducted a comparative experiment under three settings, where 4, 8, and 16 reasoning trajectories were sampled for each query, respectively. The experimental results, shown in Table~\ref{tab:grpo_sensitivity}, indicate that the overall performance of the model improves steadily as the number of sampled trajectories increases. In particular, when the number of sampled trajectories is set to 16, the model achieves the best or tied-best Macro-F1 on all four cell lines, yielding the strongest overall performance. Based on this observation, we adopt 16 sampled trajectories per query in the main experiments to obtain better overall performance.

\section{Conclusion}

In this work, we propose AROMA, an Augmented Reasoning Over a Multimodal Architecture for virtual cell genetic perturbation modeling, aiming to reliably predict gene-level expression changes under genetic perturbations while providing biologically meaningful reasoning. Comprehensive experiments show that AROMA achieves clearly higher performance than other methods, and that in zero-shot setting on unseen cell line, its accuracy degrades only mildly while remaining competitive. Stratified analyses on low-popularity and long-tail genes further indicate that AROMA maintains comparatively stable performance in knowledge-sparse regions, rather than relying on the memorization of high-frequency or highly connected genes. These experiments suggest that explicitly embedding perturbation-target relationships into a joint framework that integrates multimodal features, retrieval-grounded evidence, and a two-stage training strategy provides an effective pathway toward more reliable and interpretable virtual cell modeling. Moreover, the two knowledge graphs and the perturbation reasoning dataset constructed in this work offer reusable data foundations for future research on virtual cell domain. Looking ahead, as a general and extensible framework, AROMA will be further extended to broader perturbation settings, including combinatorial genetic perturbations and chemical interventions.

\section*{Limitations}

Although AROMA achieves stable performance gains over existing methods on genetic perturbation prediction, demonstrates good cross-cell-line generalization, and maintains reliable predictions on low-popularity and low-connectivity genes where prior knowledge is relatively sparse, this work still has three main limitations. First, the current framework is restricted to single-gene perturbation modeling and evaluation, and cannot yet directly handle more complex scenarios such as multi-gene combinational perturbations with synergistic effects or chemical interventions with diverse molecular targets. Second, AROMA focuses on three-class prediction at the level of a single target gene, where each inference outputs the expression change category for one target gene under a given perturbation and cell context, and it has not yet been extended to jointly model the perturbation effects on multiple downstream genes. Finally, the model relies on pre-constructed knowledge graphs and external textual resources to provide structural and semantic evidence during prediction, for genes that lack reliable annotations or are not covered in these databases, the available structural and functional information becomes extremely limited, which can lead to degraded prediction performance on such samples.

\section*{Ethical Considerations}

All data used in this study are derived from publicly available and legally compliant datasets and biological knowledge bases, and do not contain any personally identifiable information or clinical individual records. We carefully verified the licensing terms and usage conditions of all data sources to ensure the compliance and integrity of the research process. The knowledge graphs and perturbation reasoning data constructed in this work are released for research purposes, with the aim of supporting subsequent studies in virtual cell modeling. Furthermore, we do not anticipate any potential risks arising from this work, as it focuses on methodological contributions and is conducted entirely on non-sensitive scientific data. In addition, AI-assisted tools were used only for minor English grammar refinement and language polishing, and were not involved in any part of the research process, including study design, data processing, experimental analysis, or result interpretation.


\clearpage

\bibliography{custom}

\begin{thebibliography}{52}
\providecommand{\natexlab}[1]{#1}

\bibitem[{Adduri et~al.(2025)Adduri, Gautam, Bevilacqua, Imran, Shah, Naghipourfar, Teyssier, Ilango, Nagaraj, Dong et~al.}]{adduri2025predicting}
Abhinav~K Adduri, Dhruv Gautam, Beatrice Bevilacqua, Alishba Imran, Rohan Shah, Mohsen Naghipourfar, Noam Teyssier, Rajesh Ilango, Sanjay Nagaraj, Mingze Dong, and 1 others. 2025.
\newblock Predicting cellular responses to perturbation across diverse contexts with state.
\newblock \emph{BioRxiv}, pages 2025--06.

\bibitem[{Ahmad et~al.(2018)Ahmad, Wolberg, and Kahwaji}]{ahmad2018biochemistry}
Maria Ahmad, Adam Wolberg, and Chadi~I Kahwaji. 2018.
\newblock Biochemistry, electron transport chain.

\bibitem[{Andersson et~al.(1979)Andersson, Nilsson, and Gahmberg}]{andersson1979k562}
Leif~C Andersson, Kenneth Nilsson, and Carl~G Gahmberg. 1979.
\newblock K562—a human erythroleukemic cell line.
\newblock \emph{International journal of cancer}, 23(2):143--147.

\bibitem[{Arzumanian et~al.(2021)Arzumanian, Kiseleva, and Poverennaya}]{arzumanian2021curious}
Viktoriia~A Arzumanian, Olga~I Kiseleva, and Ekaterina~V Poverennaya. 2021.
\newblock The curious case of the hepg2 cell line: 40 years of expertise.
\newblock \emph{International journal of molecular sciences}, 22(23):13135.

\bibitem[{Ashburner et~al.(2000)Ashburner, Ball, Blake, Botstein, Butler, Cherry, Davis, Dolinski, Dwight, Eppig et~al.}]{ashburner2000gene}
Michael Ashburner, Catherine~A Ball, Judith~A Blake, David Botstein, Heather Butler, J~Michael Cherry, Allan~P Davis, Kara Dolinski, Selina~S Dwight, Janan~T Eppig, and 1 others. 2000.
\newblock Gene ontology: tool for the unification of biology.
\newblock \emph{Nature genetics}, 25(1):25--29.

\bibitem[{Bundy and Wallen(1984)}]{bundy1984breadth}
Alan Bundy and Lincoln Wallen. 1984.
\newblock Breadth-first search.
\newblock In \emph{Catalogue of artificial intelligence tools}, pages 13--13. Springer.

\bibitem[{Bunne et~al.(2024)Bunne, Roohani, Rosen, Gupta, Zhang, Roed, Alexandrov, AlQuraishi, Brennan, Burkhardt et~al.}]{bunne2024build}
Charlotte Bunne, Yusuf Roohani, Yanay Rosen, Ankit Gupta, Xikun Zhang, Marcel Roed, Theo Alexandrov, Mohammed AlQuraishi, Patricia Brennan, Daniel~B Burkhardt, and 1 others. 2024.
\newblock How to build the virtual cell with artificial intelligence: Priorities and opportunities.
\newblock \emph{Cell}, 187(25):7045--7063.

\bibitem[{Cardol et~al.(2006)Cardol, Lapaille, Minet, Franck, Matagne, and Remacle}]{cardol2006nd3}
Pierre Cardol, Marie Lapaille, Pierre Minet, Fabrice Franck, Ren{\'e}~F Matagne, and Claire Remacle. 2006.
\newblock Nd3 and nd4l subunits of mitochondrial complex i, both nucleus encoded in chlamydomonas reinhardtii, are required for activity and assembly of the enzyme.
\newblock \emph{Eukaryotic cell}, 5(9):1460--1467.

\bibitem[{Chen and Zou(2024)}]{chen2024genept}
Yiqun Chen and James Zou. 2024.
\newblock Genept: a simple but effective foundation model for genes and cells built from chatgpt.
\newblock \emph{bioRxiv}, pages 2023--10.

\bibitem[{Comanici et~al.(2025)Comanici, Bieber, Schaekermann, Pasupat, Sachdeva, Dhillon, Blistein, Ram, Zhang, Rosen et~al.}]{comanici2025gemini}
Gheorghe Comanici, Eric Bieber, Mike Schaekermann, Ice Pasupat, Noveen Sachdeva, Inderjit Dhillon, Marcel Blistein, Ori Ram, Dan Zhang, Evan Rosen, and 1 others. 2025.
\newblock Gemini 2.5: Pushing the frontier with advanced reasoning, multimodality, long context, and next generation agentic capabilities.
\newblock \emph{arXiv preprint arXiv:2507.06261}.

\bibitem[{Comerford et~al.(2014)Comerford, Huang, Du, Wang, Cai, Witkiewicz, Walters, Tantawy, Fu, Manning et~al.}]{comerford2014acetate}
Sarah~A Comerford, Zhiguang Huang, Xinlin Du, Yun Wang, Ling Cai, Agnes~K Witkiewicz, Holly Walters, Mohammed~N Tantawy, Allie Fu, H~Charles Manning, and 1 others. 2014.
\newblock Acetate dependence of tumors.
\newblock \emph{Cell}, 159(7):1591--1602.

\bibitem[{Croft et~al.(2010)Croft, O’kelly, Wu, Haw, Gillespie, Matthews, Caudy, Garapati, Gopinath, Jassal et~al.}]{croft2010reactome}
David Croft, Gavin O’kelly, Guanming Wu, Robin Haw, Marc Gillespie, Lisa Matthews, Michael Caudy, Phani Garapati, Gopal Gopinath, Bijay Jassal, and 1 others. 2010.
\newblock Reactome: a database of reactions, pathways and biological processes.
\newblock \emph{Nucleic acids research}, 39(suppl\_1):D691--D697.

\bibitem[{Cui et~al.(2024)Cui, Wang, Maan, Pang, Luo, Duan, and Wang}]{cui2024scgpt}
Haotian Cui, Chloe Wang, Hassaan Maan, Kuan Pang, Fengning Luo, Nan Duan, and Bo~Wang. 2024.
\newblock scgpt: toward building a foundation model for single-cell multi-omics using generative ai.
\newblock \emph{Nature methods}, 21(8):1470--1480.

\bibitem[{Dubey et~al.(2024)Dubey, Jauhri, Pandey, Kadian, Al-Dahle, Letman, Mathur, Schelten, Yang, Fan et~al.}]{dubey2024llama}
Abhimanyu Dubey, Abhinav Jauhri, Abhinav Pandey, Abhishek Kadian, Ahmad Al-Dahle, Aiesha Letman, Akhil Mathur, Alan Schelten, Amy Yang, Angela Fan, and 1 others. 2024.
\newblock The llama 3 herd of models.
\newblock \emph{arXiv e-prints}, pages arXiv--2407.

\bibitem[{Elnaggar et~al.(2021)Elnaggar, Heinzinger, Dallago, Rehawi, Wang, Jones, Gibbs, Feher, Angerer, Steinegger et~al.}]{elnaggar2021prottrans}
Ahmed Elnaggar, Michael Heinzinger, Christian Dallago, Ghalia Rehawi, Yu~Wang, Llion Jones, Tom Gibbs, Tamas Feher, Christoph Angerer, Martin Steinegger, and 1 others. 2021.
\newblock Prottrans: Toward understanding the language of life through self-supervised learning.
\newblock \emph{IEEE transactions on pattern analysis and machine intelligence}, 44(10):7112--7127.

\bibitem[{Fallahpour et~al.(2025)Fallahpour, Magnuson, Gupta, Ma, Naimer, Shah, Duan, Ibrahim, Goodarzi, Maddison et~al.}]{fallahpour2025bioreason}
Adibvafa Fallahpour, Andrew Magnuson, Purav Gupta, Shihao Ma, Jack Naimer, Arnav Shah, Haonan Duan, Omar Ibrahim, Hani Goodarzi, Chris~J Maddison, and 1 others. 2025.
\newblock Bioreason: Incentivizing multimodal biological reasoning within a dna-llm model.
\newblock \emph{arXiv preprint arXiv:2505.23579}.

\bibitem[{Gao et~al.(2016)Gao, Lin, Ren, Li, Chen, Yao, Yang, Jiang, Yan, Wang et~al.}]{gao2016acetate}
Xue Gao, Shu-Hai Lin, Feng Ren, Jin-Tao Li, Jia-Jia Chen, Chuan-Bo Yao, Hong-Bin Yang, Shu-Xia Jiang, Guo-Quan Yan, Di~Wang, and 1 others. 2016.
\newblock Acetate functions as an epigenetic metabolite to promote lipid synthesis under hypoxia.
\newblock \emph{Nature communications}, 7(1):11960.

\bibitem[{Gioia et~al.(2018)Gioia, Siddique, Head, Salomon, and Su}]{gioia2018genome}
Louis Gioia, Azeem Siddique, Steven~R Head, Daniel~R Salomon, and Andrew~I Su. 2018.
\newblock A genome-wide survey of mutations in the jurkat cell line.
\newblock \emph{BMC genomics}, 19(1):334.

\bibitem[{Giurgiu et~al.(2019)Giurgiu, Reinhard, Brauner, Dunger-Kaltenbach, Fobo, Frishman, Montrone, and Ruepp}]{giurgiu2019corum}
Madalina Giurgiu, Julian Reinhard, Barbara Brauner, Irmtraud Dunger-Kaltenbach, Gisela Fobo, Goar Frishman, Corinna Montrone, and Andreas Ruepp. 2019.
\newblock Corum: the comprehensive resource of mammalian protein complexes—2019.
\newblock \emph{Nucleic acids research}, 47(D1):D559--D563.

\bibitem[{{Google Cloud}(2025)}]{googlecloud2025gemini25pro}
{Google Cloud}. 2025.
\newblock Gemini 2.5 pro: Generative ai on vertex ai.
\newblock \url{https://docs.cloud.google.com/vertex-ai/generative-ai/docs/models/gemini/2-5-pro?hl=zh-cn}.
\newblock Accessed: 2025-12-21.

\bibitem[{Guo et~al.(2025)Guo, Yang, Zhang, Song, Zhang, Xu, Zhu, Ma, Wang, Bi et~al.}]{guo2025deepseek}
Daya Guo, Dejian Yang, Haowei Zhang, Junxiao Song, Ruoyu Zhang, Runxin Xu, Qihao Zhu, Shirong Ma, Peiyi Wang, Xiao Bi, and 1 others. 2025.
\newblock Deepseek-r1: Incentivizing reasoning capability in llms via reinforcement learning.
\newblock \emph{arXiv preprint arXiv:2501.12948}.

\bibitem[{Hao et~al.(2024)Hao, Gong, Zeng, Liu, Guo, Cheng, Wang, Ma, Zhang, and Song}]{hao2024large}
Minsheng Hao, Jing Gong, Xin Zeng, Chiming Liu, Yucheng Guo, Xingyi Cheng, Taifeng Wang, Jianzhu Ma, Xuegong Zhang, and Le~Song. 2024.
\newblock Large-scale foundation model on single-cell transcriptomics.
\newblock \emph{Nature methods}, 21(8):1481--1491.

\bibitem[{Hurst et~al.(2024)Hurst, Lerer, Goucher, Perelman, Ramesh, Clark, Ostrow, Welihinda, Hayes, Radford et~al.}]{hurst2024gpt}
Aaron Hurst, Adam Lerer, Adam~P Goucher, Adam Perelman, Aditya Ramesh, Aidan Clark, AJ~Ostrow, Akila Welihinda, Alan Hayes, Alec Radford, and 1 others. 2024.
\newblock Gpt-4o system card.
\newblock \emph{arXiv preprint arXiv:2410.21276}.

\bibitem[{Istrate et~al.(2025)Istrate, Milletari, Castrotorres, Tomczak, Torkar, Li, and Karaletsos}]{istrate2025rbio1}
Ana-Maria Istrate, Fausto Milletari, Fabrizio Castrotorres, Jakub~M Tomczak, Michaela Torkar, Donghui Li, and Theofanis Karaletsos. 2025.
\newblock rbio1-training scientific reasoning llms with biological world models as soft verifiers.
\newblock \emph{bioRxiv}, pages 2025--08.

\bibitem[{Ji et~al.(2023)Ji, Lee, Frieske, Yu, Su, Xu, Ishii, Bang, Madotto, and Fung}]{ji2023survey}
Ziwei Ji, Nayeon Lee, Rita Frieske, Tiezheng Yu, Dan Su, Yan Xu, Etsuko Ishii, Ye~Jin Bang, Andrea Madotto, and Pascale Fung. 2023.
\newblock Survey of hallucination in natural language generation.
\newblock \emph{ACM computing surveys}, 55(12):1--38.

\bibitem[{Karr et~al.(2012)Karr, Sanghvi, Macklin, Gutschow, Jacobs, Bolival, Assad-Garcia, Glass, and Covert}]{karr2012whole}
Jonathan~R Karr, Jayodita~C Sanghvi, Derek~N Macklin, Miriam~V Gutschow, Jared~M Jacobs, Benjamin Bolival, Nacyra Assad-Garcia, John~I Glass, and Markus~W Covert. 2012.
\newblock A whole-cell computational model predicts phenotype from genotype.
\newblock \emph{Cell}, 150(2):389--401.

\bibitem[{Kaymak et~al.(2024)Kaymak, Watson, Oswald, Ma, Johnson, DeCamp, Mabvakure, Luda, Ma, Lau et~al.}]{kaymak2024acly}
Irem Kaymak, McLane~J Watson, Brandon~M Oswald, Shixin Ma, Benjamin~K Johnson, Lisa~M DeCamp, Batsirai~M Mabvakure, Katarzyna~M Luda, Eric~H Ma, Kin Lau, and 1 others. 2024.
\newblock Acly and acss2 link nutrient-dependent chromatin accessibility to cd8 t cell effector responses.
\newblock \emph{Journal of Experimental Medicine}, 221(9):e20231820.

\bibitem[{Kim et~al.(2024)Kim, Wu, Abdulle, and Wu}]{kim2024medexqa}
Yunsoo Kim, Jinge Wu, Yusuf Abdulle, and Honghan Wu. 2024.
\newblock Medexqa: Medical question answering benchmark with multiple explanations.
\newblock \emph{arXiv preprint arXiv:2406.06331}.

\bibitem[{Klein et~al.(2025)Klein, Fleck, Bobrovskiy, Zimmermann, Becker, Palma, Dony, Tejada-Lapuerta, Huguet, Lin et~al.}]{klein2025cellflow}
Dominik Klein, Jonas~Simon Fleck, Daniil Bobrovskiy, Lea Zimmermann, S{\"o}ren Becker, Alessandro Palma, Leander Dony, Alejandro Tejada-Lapuerta, Guillaume Huguet, Hsiu-Chuan Lin, and 1 others. 2025.
\newblock Cellflow enables generative single-cell phenotype modeling with flow matching.
\newblock \emph{bioRxiv}, pages 2025--04.

\bibitem[{Lin et~al.(2023)Lin, Akin, Rao, Hie, Zhu, Lu, Smetanin, Verkuil, Kabeli, Shmueli et~al.}]{lin2023evolutionary}
Zeming Lin, Halil Akin, Roshan Rao, Brian Hie, Zhongkai Zhu, Wenting Lu, Nikita Smetanin, Robert Verkuil, Ori Kabeli, Yaniv Shmueli, and 1 others. 2023.
\newblock Evolutionary-scale prediction of atomic-level protein structure with a language model.
\newblock \emph{Science}, 379(6637):1123--1130.

\bibitem[{Ling et~al.(2022)Ling, Chen, Tang, Liu, Zhou, and Chen}]{ling2022acetyl}
Rui Ling, Gong Chen, Xiang Tang, Na~Liu, Yuepeng Zhou, and Deyu Chen. 2022.
\newblock Acetyl-coa synthetase 2 (acss2): a review with a focus on metabolism and tumor development.
\newblock \emph{Discover Oncology}, 13(1):58.

\bibitem[{Lotfollahi et~al.(2023)Lotfollahi, Klimovskaia~Susmelj, De~Donno, Hetzel, Ji, Ibarra, Srivatsan, Naghipourfar, Daza, Martin et~al.}]{lotfollahi2023predicting}
Mohammad Lotfollahi, Anna Klimovskaia~Susmelj, Carlo De~Donno, Leon Hetzel, Yuge Ji, Ignacio~L Ibarra, Sanjay~R Srivatsan, Mohsen Naghipourfar, Riza~M Daza, Beth Martin, and 1 others. 2023.
\newblock Predicting cellular responses to complex perturbations in high-throughput screens.
\newblock \emph{Molecular systems biology}, 19(6):e11517.

\bibitem[{Ma et~al.(2022)Ma, Bian, Rong, Huang, Xu, Xie, Ye, and Huang}]{ma2022cross}
Hehuan Ma, Yatao Bian, Yu~Rong, Wenbing Huang, Tingyang Xu, Weiyang Xie, Geyan Ye, and Junzhou Huang. 2022.
\newblock Cross-dependent graph neural networks for molecular property prediction.
\newblock \emph{Bioinformatics}, 38(7):2003--2009.

\bibitem[{Mering et~al.(2003)Mering, Huynen, Jaeggi, Schmidt, Bork, and Snel}]{mering2003string}
Christian~von Mering, Martijn Huynen, Daniel Jaeggi, Steffen Schmidt, Peer Bork, and Berend Snel. 2003.
\newblock String: a database of predicted functional associations between proteins.
\newblock \emph{Nucleic acids research}, 31(1):258--261.

\bibitem[{Moffett et~al.(2020)Moffett, Puthillathu, Vengilote, Jaworski, and Namboodiri}]{moffett2020acetate}
John~R Moffett, Narayanan Puthillathu, Ranjini Vengilote, Diane~M Jaworski, and Aryan~M Namboodiri. 2020.
\newblock Acetate revisited: A key biomolecule at the nexus of metabolism, epigenetics and oncogenesis—part 1: Acetyl-coa, acetogenesis and acyl-coa short-chain synthetases.
\newblock \emph{Frontiers in Physiology}, 11:580167.

\bibitem[{Montesinos-Lopez et~al.(2024)Montesinos-Lopez, Chavira-Flores, Kismiantini, Crespo-Herrera, Saint~Piere, Li, Fritsche-Neto, Al-Nowibet, Montesinos-L{\'o}pez, and Crossa}]{montesinos2024review}
Osval~A Montesinos-Lopez, Moises Chavira-Flores, Kismiantini, Leo Crespo-Herrera, Carolina Saint~Piere, HuiHui Li, Roberto Fritsche-Neto, Khalid Al-Nowibet, Abelardo Montesinos-L{\'o}pez, and Jos{\'e} Crossa. 2024.
\newblock A review of multimodal deep learning methods for genomic-enabled prediction in plant breeding.
\newblock \emph{Genetics}, 228(4):iyae161.

\bibitem[{{OpenAI}(2025)}]{openai2025o3o4systemcard}
{OpenAI}. 2025.
\newblock Openai o3 and o4-mini system card.
\newblock \url{https://openai.com/zh-Hant-HK/index/o3-o4-mini-system-card/}.
\newblock Accessed: 2025-12-21.

\bibitem[{{OpenAI Team}(2025)}]{openai2025gpt5systemcard}
{OpenAI Team}. 2025.
\newblock Gpt-5 system card.
\newblock \url{https://openai.com/index/gpt-5-system-card/}.
\newblock Accessed: 2025-12-21.

\bibitem[{Phillips et~al.(2025)Phillips, Martell, Misra, Stoisser, Prada-Medina, Donovan-Maiye, and M{\"a}rtens}]{phillips2025synthpert}
Lawrence Phillips, Marc~Boubnovski Martell, Aditya Misra, Josefa~Lia Stoisser, Cesar~A Prada-Medina, Rory Donovan-Maiye, and Kaspar M{\"a}rtens. 2025.
\newblock Synthpert: Enhancing llm biological reasoning via synthetic reasoning traces for cellular perturbation prediction.
\newblock \emph{arXiv preprint arXiv:2509.25346}.

\bibitem[{Pryor et~al.(2006)Pryor, Houk, Foote, Fukuto, Ignarro, Squadrito, and Davies}]{pryor2006free}
William~A Pryor, Kendall~N Houk, Christopher~S Foote, Jon~M Fukuto, Louis~J Ignarro, Giuseppe~L Squadrito, and Kelvin~JA Davies. 2006.
\newblock Free radical biology and medicine: it's a gas, man!
\newblock \emph{American Journal of Physiology-Regulatory, Integrative and Comparative Physiology}, 291(3):R491--R511.

\bibitem[{Roohani et~al.(2024)Roohani, Huang, and Leskovec}]{roohani2024predicting}
Yusuf Roohani, Kexin Huang, and Jure Leskovec. 2024.
\newblock Predicting transcriptional outcomes of novel multigene perturbations with gears.
\newblock \emph{Nature Biotechnology}, 42(6):927--935.

\bibitem[{Sharma et~al.(2009)Sharma, Lu, and Bai}]{sharma2009mitochondrial}
Lokendra~K Sharma, Jianxin Lu, and Yidong Bai. 2009.
\newblock Mitochondrial respiratory complex i: structure, function and implication in human diseases.
\newblock \emph{Current medicinal chemistry}, 16(10):1266--1277.

\bibitem[{Sharrow et~al.(2024)Sharrow, Megill, Chen, Farooqi, Tangudu, Uboveja, McGonigal, Hempel, Snyder, Buckanovich et~al.}]{sharrow2024acetate}
Allison~C Sharrow, Emily Megill, Amanda~J Chen, Afifa Farooqi, Naveen~Kumar Tangudu, Apoorva Uboveja, Stacy McGonigal, Nadine Hempel, Nathaniel~W Snyder, Ronald~J Buckanovich, and 1 others. 2024.
\newblock Acetate drives ovarian cancer quiescence via acss2-mediated acetyl-coa production.
\newblock \emph{Molecular Metabolism}, 89:102031.

\bibitem[{Spalluto et~al.(2013)Spalluto, Wilson, and Hearn}]{spalluto2013evidence}
Cosma Spalluto, David~I Wilson, and Tom Hearn. 2013.
\newblock Evidence for reciliation of rpe1 cells in late g1 phase, and ciliary localisation of cyclin b1.
\newblock \emph{FEBS open bio}, 3:334--340.

\bibitem[{Veli{\v{c}}kovi{\'c} et~al.(2017)Veli{\v{c}}kovi{\'c}, Cucurull, Casanova, Romero, Lio, and Bengio}]{velivckovic2017graph}
Petar Veli{\v{c}}kovi{\'c}, Guillem Cucurull, Arantxa Casanova, Adriana Romero, Pietro Lio, and Yoshua Bengio. 2017.
\newblock Graph attention networks.
\newblock \emph{arXiv preprint arXiv:1710.10903}.

\bibitem[{Wei et~al.(2025)Wei, Ma, Wang, Li, Song, and Zheng}]{wei2025vcworld}
Zhijian Wei, Runze Ma, Zichen Wang, Zhongmin Li, Shuotong Song, and Shuangjia Zheng. 2025.
\newblock Vcworld: A biological world model for virtual cell simulation.
\newblock \emph{arXiv preprint arXiv:2512.00306}.

\bibitem[{White(2020)}]{white2020pubmed}
Jacob White. 2020.
\newblock Pubmed 2.0.
\newblock \emph{Medical reference services quarterly}, 39(4):382--387.

\bibitem[{Wu et~al.(2025)Wu, Littman, Levine, Qiu, Biancalani, Richmond, and Huetter}]{wu2025contextualizing}
Menghua Wu, Russell Littman, Jacob Levine, Lin Qiu, Tommaso Biancalani, David Richmond, and Jan-Christian Huetter. 2025.
\newblock Contextualizing biological perturbation experiments through language.
\newblock \emph{arXiv preprint arXiv:2502.21290}.

\bibitem[{Yang et~al.(2025)Yang, Li, Yang, Zhang, Hui, Zheng, Yu, Gao, Huang, Lv et~al.}]{yang2025qwen3}
An~Yang, Anfeng Li, Baosong Yang, Beichen Zhang, Binyuan Hui, Bo~Zheng, Bowen Yu, Chang Gao, Chengen Huang, Chenxu Lv, and 1 others. 2025.
\newblock Qwen3 technical report.
\newblock \emph{arXiv preprint arXiv:2505.09388}.

\bibitem[{Ye et~al.(2025)Ye, Cai, Lai, Wang, Huang, Wang, Liu, and Zeng}]{ye2025drugassist}
Geyan Ye, Xibao Cai, Houtim Lai, Xing Wang, Junhong Huang, Longyue Wang, Wei Liu, and Xiangxiang Zeng. 2025.
\newblock Drugassist: A large language model for molecule optimization.
\newblock \emph{Briefings in Bioinformatics}, 26(1):bbae693.

\bibitem[{Yuan et~al.(2025)Yuan, Li, Ye, Zhang, Huang, Huang, Liu, Yao, and Rong}]{yuan2025annotation}
Chaohao Yuan, Songyou Li, Geyan Ye, Yikun Zhang, Long-Kai Huang, Wenbing Huang, Wei Liu, Jianhua Yao, and Yu~Rong. 2025.
\newblock Annotation-guided protein design with multi-level domain alignment.
\newblock In \emph{Proceedings of the 31st ACM SIGKDD Conference on Knowledge Discovery and Data Mining V. 1}, pages 1855--1866.

\bibitem[{Zhang et~al.(2025)Zhang, Ye, Yuan, Han, Huang, Yao, Liu, and Rong}]{zhang2025atomas}
Yikun Zhang, Geyan Ye, Chaohao Yuan, Bo~Han, Long-Kai Huang, Jianhua Yao, Wei Liu, and Yu~Rong. 2025.
\newblock Atomas: Hierarchical adaptive alignment on molecule-text for unified molecule understanding and generation.
\newblock In \emph{The Thirteenth International Conference on Learning Representations}.

\end{thebibliography}

\appendix

\section{Algorithmic Framework of AROMA}

This section provides a formal overview of the algorithmic framework of AROMA. Algorithm 1 presents a procedural description of the overall model, covering the collaborative workflow among retrieval-augmented contextualization, multimodal interaction feature injection, and the two-stage training strategy.

\label{sec:appendix}

\begin{algorithm}[t!]
\caption{AROMA Multimodal Training Framework}
\label{alg:aroma_training}
\begin{algorithmic}[1]
    \Require 
        Multimodal reasoning dataset $\mathcal{D}_{\text{sft}}$; 
        Pre-trained encoders $\{E_{\text{gene}}, E_{\text{path}}, E_{\text{seq}}\}$; 
        Base LLM parameters $\Theta_{\text{LLM}}$.
    \Ensure 
        Optimized parameters $\Theta^*$ (LLM via LoRA) and $\Phi^*$ (Interaction \& Projectors).

    \State Initialize interaction modules $\Phi = \{\text{CrossAttn}_m, \Phi^{m}_{\text{proj}}\}_{m \in \{\text{gene, path, prot}\}}$
    \State Attach LoRA adapters to $\Theta_{\text{LLM}}$; Freeze encoders $\{E_{\text{gene}}, E_{\text{path}}, E_{\text{seq}}\}$

    \While{not converged}
        \State Sample mini-batch $\mathcal{B} = \{(q, Y_{\text{reason}}, Y_{\text{answer}})\} \sim \mathcal{D}_{\text{sft}}$
        \State Parse perturbation gene $g_p$ and target gene $g_t$ from query $q$

        \Statex \textbf{Step 1: Retrieval-Augmented Contextualization (Sec 3.3)}
        \State \textsc{RetrieveContext}($T_{\text{desc}}, T_{\text{path}}, T_{\text{cell}}$) for $g_p$ and $g_t$
        \State $X_{\text{text}} \leftarrow \Phi(q, T_{\text{desc}}, T_{\text{path}}, T_{\text{cell}})$ 

        \Statex \textbf{Step 2: Multimodal Interaction Encoding (Sec 3.4)}
        \State $H^{(p)}_{m}, H^{(t)}_{m} \leftarrow E_m(g_p, g_t)$
        \For{$m \in \{\text{gene, path, protein}\}$}
            \State $z^{m}_{\text{inter}} \leftarrow \textsc{CrossAttn}_m(H^{(p)}_{m}, H^{(t)}_{m})$ 
            \State $e^{m}_{\text{inter}} \leftarrow \Phi^{m}_{\text{proj}}(z^{m}_{\text{inter}})$
        \EndFor
        \State $X \leftarrow \textsc{Inject}(X_{\text{text}}, \{e^{m}_{\text{inter}}\}_m)$ \Comment{Combine text and multimodal tokens}

        \Statex \textbf{Step 3: Two-Stage Optimization (Sec 3.5)}
        \If{Stage is \textbf{SFT}}
            \State $\mathcal{L}_{\text{SFT}} \leftarrow -\sum \log P_{\Theta}(y_t \mid y_{<t}, X)$
            \State \textbf{Update:} $\Theta_{\text{LLM}}$ (LoRA) and $\Phi$; \textbf{Freeze:} $\{E_{\text{gene}}, E_{\text{path}}, E_{\text{seq}}\}$
        \ElsIf{Stage is \textbf{GRPO}}
            \State Sample trajectories $\tau$; Compute rewards $R(\tau)$ based on accuracy and format
            \State $\mathcal{L}_{\text{GRPO}} \leftarrow \mathbb{E}_{\tau \sim \pi_{\Theta}} [R(\tau) - \beta \text{KL}(\pi_{\Theta} \| \pi_{\text{ref}})]$
            \State \textbf{Update:} $\Theta_{\text{LLM}}$ (LoRA) \textit{only}; \textbf{Freeze:} all other modules ($\Phi$ and $E$)
        \EndIf 
    \EndWhile
    \State \Return $\Theta^*, \Phi^*$
\end{algorithmic}
\end{algorithm}

\section{Data Processing}

\subsection{Ternary Label Construction}

We normalized raw single-cell gene expression counts using a log-transformed counts-per-ten-thousand scheme. For a cell $i$ and gene $j$ with raw UMI count $c_{ij}$, the normalized expression value is computed as:
\begin{equation}
\small
x_{ij} = \log\left(\frac{c_{ij}}{\sum_{j} c_{ij}} \cdot 10000 + 1\right)
\end{equation}
where $x_{ij}$ denotes the normalized expression of gene $j$ in cell $i$.
For each perturbation and target gene, we compare perturbation cells with the corresponding non-targeting control (NTC) cells by performing a Wilcoxon signed-rank test, and apply the Benjamini--Hochberg procedure to obtain an adjusted $p$-value $\tilde{p}$. In parallel, we compute the difference in mean normalized expression between perturbation and control cells for each target gene $j$:
\begin{equation}
\small
\Delta \mu_j = \mathrm{mean}_{\text{perturbation}}(x_{j}) - \mathrm{mean}_{\text{NTC}}(x_{j})
\end{equation}
where $x_{j}$ denotes the vector of normalized expression values of gene $j$ across cells, and $\Delta \mu_j$ is the corresponding difference in mean expression between the perturbation and NTC groups.
Based on $\tilde{p}$ and the sign of $\Delta \mu_j$, each cell line-perturbation-target gene triplet is mapped to a three-way label: if $\tilde{p} < 0.05$ and $\Delta \mu_j > 0$, the sample is labeled as upregulated (Up); if $\tilde{p} < 0.05$ and $\Delta \mu_j < 0$, it is labeled as downregulated (Down); all remaining cases are labeled as non-differentially expressed (ND). This yields a ternary classification over \{Up, Down, ND\} for downstream modeling. The corresponding class counts for each cell line are shown in Table~\ref{tab:dataset_stats}.

\begin{table}[H]
\centering
\caption{Summary of samples for each cell line.}
\label{tab:dataset_stats}
\resizebox{\columnwidth}{!}{%
\begin{tabular}{l l r r r r}
\specialrule{1.5pt}{0pt}{0pt}
Cell line & Split & Total & ND & Up & Down \\
\midrule
K562   & Train & 134,467 & 117,606 & 11,041 & 5,820 \\
       & Test  &  23,212 &  20,093 &  2,530 &   589 \\
\midrule
HepG2  & Train & 101,140 &  86,883 &  6,249 &  8,008 \\
       & Test  &  25,749 &  22,146 &  1,599 &  2,004 \\
\midrule
Jurkat & Train & 113,684 &  97,747 &  5,119 & 10,818 \\
       & Test  &  29,138 &  25,017 &  1,379 &  2,742 \\
\midrule
RPE1   & Train & 149,147 & 127,860 &  8,381 & 12,906 \\
       & Test  &  37,942 &  32,577 &  2,121 &  3,244 \\
\specialrule{1.5pt}{0pt}{0pt}
\end{tabular}%
}
\end{table}

\subsection{Sample Construction for GNN Encoders Pre-training}
\label{app:gnnsample}
For GNN encoders pre-training on both the Gene-KG and Path-KG, we formulate the task as a link prediction problem and construct samples based on the connectivity patterns of the knowledge graphs. Concretely, all observed edges between nodes in a graph are treated as \textit{positive} samples. In parallel, \textit{negative} samples are generated by sampling node pairs that are not connected in the graph, and the overall number of negative samples is controlled such that the ratio between positive and negative samples is kept at 1:1, yielding a balanced binary classification dataset for link prediction. We split all constructed samples into training, validation, and test sets with proportions of 80\%, 10\%, and 10\%, respectively. The training set is used for learning model parameters, the validation set is used for monitoring and evaluating model performance during training, and the test set is used for evaluating the final trained model. The above sample construction and data splitting procedure is applied independently to each of the two knowledge graphs, resulting in two separate link prediction sample sets that are used to pre-train the GNN encoders on the Gene-KG and Path-KG, respectively. The numbers of positive and negative samples constructed for pre-training on the two knowledge graphs are summarized in Table~\ref{tab:gnn_pretrain_samples}.

\begin{table}[H]
\centering
\caption{Summary of positive and negative samples for GNN encoders pre-training on the two knowledge graphs.}
\label{tab:gnn_pretrain_samples}
\resizebox{\columnwidth}{!}{%
\begin{tabular}{l l r r r}
\specialrule{1.5pt}{0pt}{0pt}
Graph   & Split      & Positive & Negative & Total \\
\midrule
\multirow{4}{*}{Gene-KG} 
        & Train      & 602,090 & 602,090 & 1,204,180 \\
        & Validation &  75,261 &  75,261 &   150,522 \\
        & Test       &  75,261 &  75,261 &   150,522 \\
        & Total      & 752,612 & 752,612 & 1,505,224 \\
\midrule
\multirow{4}{*}{Path-KG} 
        & Train      & 352,941 & 352,941 &   705,882 \\
        & Validation &  44,118 &  44,118 &    88,236 \\
        & Test       &  44,117 &  44,117 &    88,234 \\
        & Total      & 441,176 & 441,176 &   882,352 \\
\specialrule{1.5pt}{0pt}{0pt}
\end{tabular}%
}
\end{table}

\section{Supplementary Experimental Results}
\label{app:supp_exp}

\subsection{Effect of Local Subgraph Range}
\label{app:subgraph_range}

\begin{table*}[t]
\centering
\small
\caption{F1-scores for different $k$-hop local subgraphs across cell lines.}
\label{tab:khop_results}
\begin{tabular*}{\textwidth}{@{\extracolsep{\fill}}lcccccccccccc}
\toprule
\multirow{2}{*}{Subgraph Range} &
\multicolumn{3}{c}{K562} &
\multicolumn{3}{c}{HepG2} &
\multicolumn{3}{c}{Jurkat} &
\multicolumn{3}{c}{RPE1 (Zero-shot)} \\
\cmidrule(lr){2-4}
\cmidrule(lr){5-7}
\cmidrule(lr){8-10}
\cmidrule(lr){11-13}
& ND & Up & Down & ND & Up & Down & ND & Up & Down & ND & Up & Down \\
\midrule
1-hop & \textbf{0.97} & 0.67 & 0.39 & \textbf{0.97} & 0.66 & 0.69 & \textbf{0.97} & \textbf{0.57} & 0.69 & \textbf{0.96} & \textbf{0.57} & \textbf{0.66} \\
2-hop & 0.96 & \textbf{0.69} & 0.37 & \textbf{0.97} & 0.63 & \textbf{0.70} & \textbf{0.97} & 0.56 & \textbf{0.70} & \textbf{0.96} & 0.55 & \textbf{0.66} \\
3-hop & \textbf{0.97} & 0.66 & \textbf{0.40} & \textbf{0.97} & \textbf{0.67} & 0.69 & 0.96 & 0.55 & 0.69 & \textbf{0.96} & 0.56 & 0.65 \\
\bottomrule
\end{tabular*}
\end{table*}

In the multimodal interaction encoding stage, AROMA extracts local $k$-hop subgraphs around both the perturbation gene and the target gene from the Gene-KG and the Path-KG, and encodes these subgraphs with the pretrained GNN encoders followed by mean pooling to obtain structural embeddings. To systematically assess how the neighborhood range affects predictive performance, we vary $k \in \{1,2,3\}$ when constructing these local subgraphs. For each setting, we train and evaluate the model on the K562, HepG2, and Jurkat cell lines, and additionally perform unseen-cell-line zero-shot evaluation on RPE1. The experimental results are shown in Table~\ref{tab:khop_results}.
The results indicate that the F1-scores obtained with 1-hop, 2-hop, and 3-hop subgraphs are overall very similar across cell lines, indicating that AROMA is relatively robust to the choice of local neighborhood size. Nevertheless, 1-hop subgraphs yield slightly better performance in most cases. Consequently, all main-text experiments adopt 1-hop local subgraphs as the default configuration for structural evidence extraction.

\subsection{Base Model Selection}
\label{app:base_model}

To select an appropriate base language model as the backbone of AROMA, we compare three candidate models: Deepseek-Distilled-Llama-8B, Llama3-8B, and Qwen3-8B. For each candidate, we perform supervised fine-tuning on the domain-specific reasoning dataset, train and evaluate the model on the K562, HepG2, and Jurkat cell lines, and additionally conduct zero-shot evaluation on the unseen RPE1 cell line. The results are summarized in Table~\ref{tab:base_model}. The experimental results show that using Qwen3-8B as the base model yields the best overall performance, with higher classification metrics and a better average score across cell lines compared to the other candidates. Therefore, we adopt Qwen3-8B as the base language model of AROMA and use it as the default backbone in all subsequent experiments.

\subsection{Experimental Results of GNN Pre-training}
\label{app:gnn_pretrain}

The two GNN encoders are pre-trained on the Gene-KG and Path-KG using a binary edge-existence prediction objective, and AUROC is used as the evaluation metric. The experimental results are reported in Table~\ref{tab:gnn_pretrain_auroc}, which show that the learned topological representations exhibit good discriminative capacity and provide a reliable structural basis for the subsequent multimodal interaction modeling.

\begin{table}[H]
\centering
\small
\caption{AUROC of GNN pre-training on the two knowledge graphs.}
\label{tab:gnn_pretrain_auroc}
\begin{tabular*}{\linewidth}{@{\extracolsep{\fill}}lc}
\toprule
Knowledge graph & Test AUROC \\
\midrule
Gene-KG  & 0.95 \\
Path-KG  & 0.92 \\
\bottomrule
\end{tabular*}
\end{table}

\subsection{Stratified Robustness Results under Knowledge Sparsity}
\label{app:robustness_stratified}

In Section~\ref{subsec:knowledge sparsity}, we conduct a stratified robustness evaluation of AROMA under different gene popularity and node degree groups. The complete results are reported in Table~\ref{tab:all_cells_pop_degree}.

\begin{table}[H]
\centering
\small
\setlength{\tabcolsep}{4pt}
\caption{Stratified robustness evaluation on all cell lines by
gene popularity and node degree.}
\begin{tabular}{lllccc}
\toprule
Cell & Model & Group & High & Mid & Low \\
\midrule
\multirow{4}{*}{K562} 
& \multirow{2}{*}{w/o R\&M} 
    & Gene Popularity & 0.65 & 0.57 & 0.53 \\
&   & Node Degree    & 0.65 & 0.59 & 0.51 \\
& \multirow{2}{*}{AROMA} 
    & Gene Popularity & 0.69 & 0.64 & 0.61 \\
&   & Node Degree    & 0.71 & 0.62 & 0.61 \\
\midrule
\multirow{4}{*}{HepG2} 
& \multirow{2}{*}{w/o R\&M} 
    & Gene Popularity & 0.76 & 0.63 & 0.57 \\
&   & Node Degree    & 0.73 & 0.65 & 0.59 \\
& \multirow{2}{*}{AROMA} 
    & Gene Popularity & 0.78 & 0.69 & 0.67 \\
&   & Node Degree    & 0.79 & 0.68 & 0.65 \\
\midrule
\multirow{4}{*}{Jurkat} 
& \multirow{2}{*}{w/o R\&M} 
    & Gene Popularity & 0.71 & 0.65 & 0.59 \\
&   & Node Degree    & 0.69 & 0.67 & 0.61 \\
& \multirow{2}{*}{AROMA} 
    & Gene Popularity & 0.75 & 0.72 & 0.67 \\
&   & Node Degree    & 0.73 & 0.72 & 0.68 \\
\midrule
\multirow{4}{*}{RPE1} 
& \multirow{2}{*}{w/o R\&M} 
    & Gene Popularity & 0.69 & 0.62 & 0.56 \\
&   & Node Degree    & 0.72 & 0.63 & 0.57 \\
& \multirow{2}{*}{AROMA} 
    & Gene Popularity & 0.77 & 0.74 & 0.71 \\
&   & Node Degree    & 0.79 & 0.75 & 0.69 \\
\bottomrule
\end{tabular}
\label{tab:all_cells_pop_degree}
\end{table}

\begin{table*}[t]
\centering
\small
\setlength{\tabcolsep}{3pt}
\caption{Comparison of different base language models on the genetic perturbation prediction. All models are fine-tuned on the domain-specific reasoning dataset. Reported values are F1-scores for Non-differential (ND), Up-regulated (Up), and Down-regulated (Down) predictions across cell lines; RPE1 is evaluated in the unseen-cell-line zero-shot setting.}
\label{tab:base_model}
\begin{tabular*}{\textwidth}{@{\extracolsep{\fill}}lcccccccccccccc}
\toprule
\multirow{2}{*}{Model} &
\multicolumn{3}{c}{K562} &
\multicolumn{3}{c}{HepG2} &
\multicolumn{3}{c}{Jurkat} &
\multicolumn{3}{c}{RPE1 (Zero-shot)} &
\multirow{2}{*}{Average} \\
\cmidrule(lr){2-4}
\cmidrule(lr){5-7}
\cmidrule(lr){8-10}
\cmidrule(lr){11-13}
& ND & Up & Down & ND & Up & Down & ND & Up & Down & ND & Up & Down & \\
\midrule
Deepseek-Distilled-Llama-8B (SFT)
& \textbf{0.94} & 0.56 & \textbf{0.31}
& \textbf{0.95} & 0.52 & \textbf{0.57}
& 0.95 & 0.41 & \textbf{0.60}
& 0.94 & 0.42 & 0.53
& 0.64 \\

Llama3-8B (SFT)
& \textbf{0.94} & 0.51 & 0.27
& \textbf{0.95} & 0.47 & 0.56
& 0.95 & 0.38 & 0.53
& 0.94 & 0.36 & 0.48
& 0.61 \\

Qwen3-8B (SFT)
& \textbf{0.94} & \textbf{0.59} & \textbf{0.31}
& \textbf{0.95} & \textbf{0.53} & 0.55
& \textbf{0.96} & \textbf{0.43} & 0.57
& \textbf{0.95} & \textbf{0.45} & \textbf{0.54}
& \textbf{0.65} \\
\bottomrule
\end{tabular*}
\end{table*}

\section{Experimental Setup}
\label{app:case_study2}

\subsection{Model Architecture}
\label{app:model_architecture}
\noindent\textbf{Graph Neural Network Encoders.}
In the structural modality, all graph encoders used in this work adopt a unified 
Graph Attention Network architecture. The numbers of layers and the corresponding
hyperparameter configurations are summarized in Table~\ref{tab:gat_arch}.

\begin{table}[H]
\centering
\small
\caption{Architecture configuration of the pretrained GAT encoders.}
\label{tab:gat_arch}
\begin{tabular*}{\linewidth}{@{}l@{\extracolsep{\fill}}l@{}}
\toprule
Component & Setting \\
\midrule
Input feature dimension    & 1024 \\
Hidden dimension           & 64 \\
Output embedding dimension & 1024 \\
Number of layers           & 3 \\
Attention heads            & 4 \\
Activation function        & ELU \\
Dropout rate               & 0.5 \\
\bottomrule
\end{tabular*}
\end{table}

\noindent\textbf{Protein Sequence Encoder.}
For the sequence modality, we use the pretrained ESM-2 model, and adopt the
\texttt{esm2\_t48\_15B\_UR50D} variant in all experiments.

\subsection{Training Hyperparameters}
\label{app:training_hyperparameters}

\noindent\textbf{GNN Pre-Training.}
For structural representation learning on the Gene-KG and Path-KG, 
we pre-train the GNN encoders under a link prediction objective. 
The core training hyperparameters used in the GNN pre-training stage 
are summarized in Table~\ref{tab:gnn_pretrain_hparams}.

\begin{table}[H]
\centering
\small
\caption{Hyperparameters used in GNN pre-training.}
\label{tab:gnn_pretrain_hparams}
\begin{tabular*}{\linewidth}{@{}l@{\extracolsep{\fill}}l@{}}
\toprule
Hyperparameter & Value \\
\midrule
Optimizer & AdamW \\
Learning rate & 1e-3 \\
Training epochs & 200 \\
\bottomrule
\end{tabular*}
\end{table}

\noindent\textbf{Multimodal Supervised Fine-Tuning.} 
In the two-stage training pipeline, we construct the GRPO training set by randomly sampling 5000 training instances from each cell-line subset of the full training corpus. The sampled instances are used exclusively for the second-stage GRPO optimization, while the remaining training samples are used for the first-stage multimodal supervised fine-tuning. The main training hyperparameters used in the multimodal supervised fine-tuning stage are summarized in Table~\ref{tab:sft_hparams}.

\begin{table}[H]
\centering
\small
\caption{Hyperparameters used in multimodal supervised fine-tuning.}
\label{tab:sft_hparams}
\begin{tabular*}{\linewidth}{@{}l@{\extracolsep{\fill}}l@{}}
\toprule
Hyperparameter & Value \\
\midrule
Base LLM & Qwen3-8B \\
LoRA rank $r$ & 16 \\
LoRA $\alpha$ & 32 \\
LoRA dropout & 0.1 \\
Target modules & All linear layers \\
Per-device train batch size & 4 \\
Gradient accumulation steps & 4 \\
Learning rate & 1e-4 \\
Training epochs & 3 \\
Precision & bfloat16 \\
\bottomrule
\end{tabular*}
\end{table}

\noindent\textbf{GRPO Reinforcement Learning.}
The main training hyperparameters used in the GRPO stage are summarized in Table~\ref{tab:grpo_hparams}.

\begin{table}[H]
\centering
\small
\caption{Hyperparameters used in GRPO reinforcement learning.}
\label{tab:grpo_hparams}
\begin{tabular*}{\linewidth}{@{}l@{\extracolsep{\fill}}l@{}}
\toprule
Hyperparameter & Value \\
\midrule
Learning rate & 1e-6 \\
Adam $\beta_1$ & 0.9 \\
Adam $\beta_2$ & 0.99 \\
Weight decay & 0.1 \\
Warmup ratio & 0.1 \\
LR scheduler & Cosine \\
Batch size (per device) & 2 \\
Gradient accumulation steps & 8 \\
Number of generations per query & 16 \\
Max completion length & 2048 \\
Training epochs & 2 \\
Precision & bfloat16 \\
Temperature & 1.2 \\
Top-p & 0.9 \\
\bottomrule
\end{tabular*}
\end{table}

\subsection{Baseline Configurations}
\label{app:baseline_config}

\noindent\textbf{General LLMs}
For the general-purpose large language model baselines reported in
Table~\ref{tab:performance_merged}, we use the official API versions listed in
Table~\ref{tab:baseline_general_llms}.

\begin{table}[t!]
\centering
\small
\caption{Model versions used for the General LLM baselines.}
\label{tab:baseline_general_llms}
\begin{tabular*}{\linewidth}{@{}l@{\extracolsep{\fill}}l@{}}
\toprule
\textbf{Model} & \textbf{Version} \\
\midrule
DeepSeek-R1 & \texttt{DeepSeek-R1-0528} \\
OpenAI o4-mini & \texttt{openai-o4-mini-2025-04-16} \\
GPT-5 & \texttt{gpt-5-2025-08-07} \\
Gemini-2.5-pro & \texttt{gemini-2.5-pro-2025-06} \\
Qwen3-235B & \texttt{Qwen3-235B-A22B-Instruct-2507} \\
\bottomrule
\end{tabular*}
\end{table}

\noindent\textbf{Domain-Specific Foundation Models}
For the domain-specific foundation model baselines reported in Table~\ref{tab:performance_merged}, we strictly follow the official implementations and training procedures of each method when reproducing and evaluating the results. All models are trained and evaluated using the standard PerturbQA training-test splits, and we adopt the best hyperparameter settings reported by the original authors.

As the models used in this study are computationally expensive and a single full training run typically requires more than 30 hours, all experimental results reported in this paper are obtained from a single full training and evaluation run under the publicly released hyperparameter configuration.

\section{Case Study}
\label{sec:case_study}

\subsection{Reasoning Pipeline and Source Tracing}

\begin{figure*}[t]
    \centering
    \includegraphics[width=\textwidth]{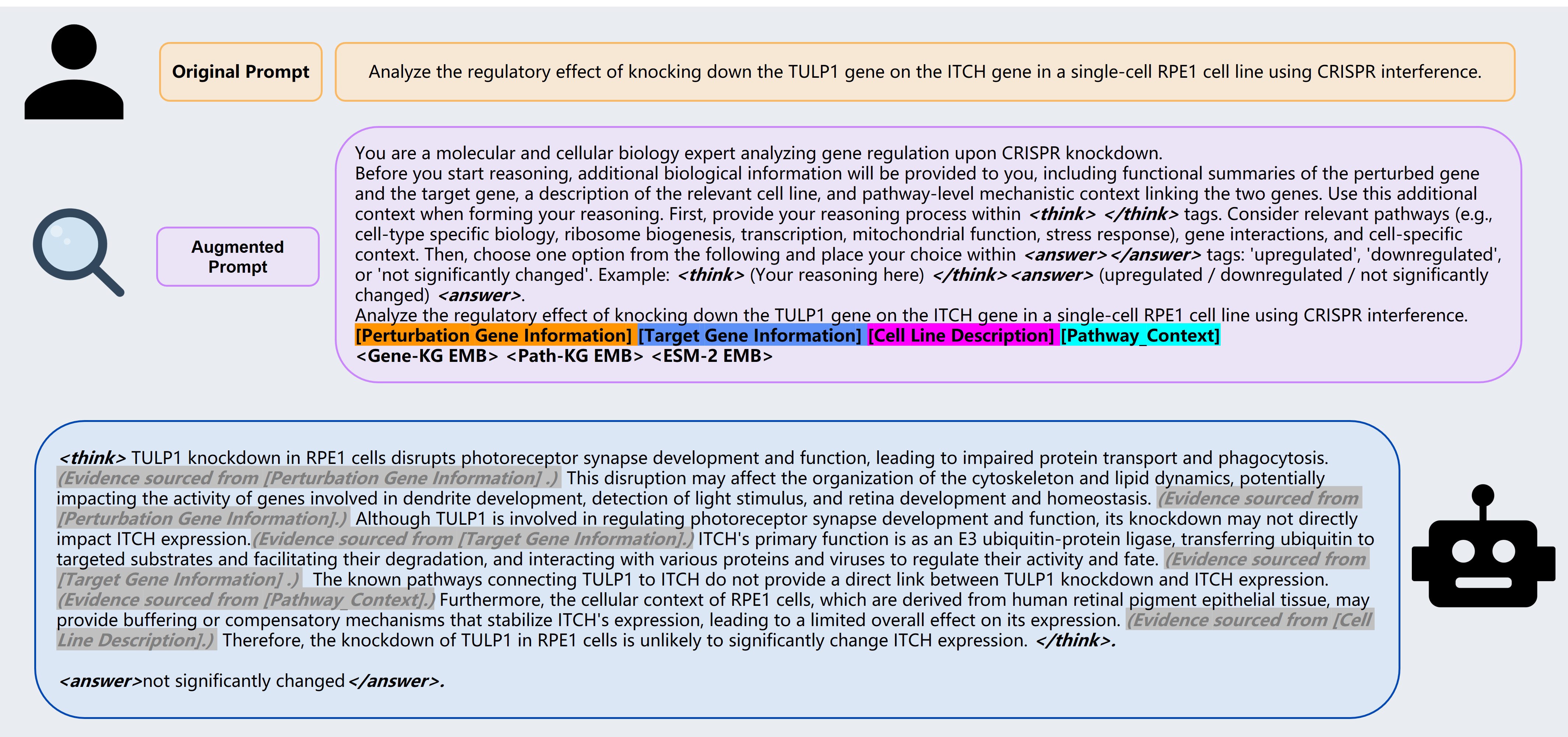}
    \caption{Visualization of the sentence-level source-tracing analysis.}
    \label{fig:case_study}
\end{figure*}

In this section, we present a case study to illustrate the complete reasoning and generation pipeline of AROMA, including the process from the initial query, to constructing a retrieval-augmented prompt based on biological evidence, and finally to generating the model's reasoning output. In addition, we perform sentence-level source-tracing analysis on the generated answer to verify that each key statement can be grounded in the retrieved biological evidence, thereby demonstrating that the model output is supported by explicit knowledge sources. A visualization of the sentence-level source-tracing analysis is shown in Figure~\ref{fig:case_study}.

At first, the model receives the following original query:

\vspace{0.15cm}
\noindent\textbf{Example of Original Query.}
\vspace{0.15cm}

\noindent
{\itshape
"Analyze the regulatory effect of knocking down the TULP1 gene on the ITCH gene in a single-cell RPE1 cell line using CRISPR interference."
}
\vspace{0.15cm}

After receiving the original query, AROMA automatically retrieves biologically relevant evidence from the Pathway Knowledge Graph and external biological databases. The retrieved evidence includes functional descriptions of the perturbation and target genes, pathway-level regulatory connections between them, and cell-line-specific contextual information. An example of the retrieved information is shown below.

\vspace{0.15cm}
\noindent\textbf{Example of the retrieved information.}
\vspace{0.15cm}

\noindent
{\itshape
"\textbf{Perturbation\_gene}": "TULP1" \\
"\textbf{Target\_gene}": "ITCH" \\
"\textbf{Cell\_line}": "RPE1" \\
"\textbf{Perturbation\_Gene\_summary\_1}": "Perturbing TULP1 via gene knockdown disrupts photoreceptor synapse development and function, leading to impaired protein transport and phagocytosis in photoreceptor cells. This knockdown also affects the long-term survival of photoreceptor cells, potentially impacting eye function. Furthermore, the loss of TULP1 function may alter the organization of the cytoskeleton and lipid dynamics, given its interactions with these components. Additionally, the knockdown may influence the activity of genes involved in dendrite development, detection of light stimulus, and retina development and homeostasis, such as ARHGAP26, NRXN1, and TULP3. Overall, perturbing TULP1 may have far-reaching consequences for photoreceptor cell biology and visual perception." \\
"\textbf{Perturbation\_Gene\_summary\_2}": "The primary molecular and cellular function of gene TULP1 is to regulate photoreceptor synapse development and function, as well as facilitate protein transport and phagocytosis in photoreceptor cells, through its interactions with cytoskeleton proteins and lipids. Additionally, TULP1 contributes to the long-term survival of photoreceptor cells. Perturbing gene TULP1 via gene knockdown may disrupt normal photoreceptor function and development, leading to impaired photoreceptor cell survival and potentially affecting eye function." \\
"\textbf{Target\_Gene\_summary\_1}": "Perturbations that may impact the levels of ITCH include alterations in the activity of upstream regulators, such as SMAD3, or changes in the availability of its binding partners, including viral proteins and other cellular substrates. Additionally, perturbations in the formation or stability of protein complexes that ITCH participates in, such as the AIP4-DTX3L complex or the CPSF6-EWSR1-ITCH-NUDT21-POLR2A-UBAP2L complex, may also affect ITCH levels. Furthermore, changes in the ubiquitination or deubiquitination of ITCH itself, or its substrates, could also influence ITCH levels." \\
"\textbf{Target\_gene\_summary\_2}": "The primary molecular and cellular function of gene ITCH is to act as an E3 ubiquitin-protein ligase, transferring ubiquitin to targeted substrates and facilitating their degradation, and to interact with various proteins and viruses to regulate their activity and fate. ITCH also participates in the formation of protein complexes that mediate ubiquitination and proteasomal degradation of specific targets. Perturbations that might impact the expression of gene ITCH include alterations in the activity of upstream regulators, such as SMAD3, or changes in the availability of its binding partners, including viral proteins and other cellular substrates." \\
"\textbf{Pathway\_text}": "Pathway 1: TULP1 $\rightarrow$ photoreceptor inner segment $\rightarrow$ DRAM2 $\rightarrow$ apoptotic process $\rightarrow$ ITCH. Pathway 2: TULP1 $\rightarrow$ phosphatidylinositol-4,5-bisphosphate binding $\rightarrow$ PFN1 $\rightarrow$ cell cortex $\rightarrow$ ITCH. Pathway 3: TULP1 $\rightarrow$ photoreceptor inner segment $\rightarrow$ SAG $\rightarrow$ G protein-coupled receptor internalization $\rightarrow$ receptor internalization $\rightarrow$ ITCH." \\
"\textbf{Cell\_line\_description}": "The RPE1 cell line is derived from human retinal pigment epithelial tissue, a monolayer of hexagonal, melanin-containing cells that support photoreceptor function. These cells absorb stray light, maintain the blood-retina barrier, regulate ion and nutrient transport, recycle visual cycle components, phagocytose photoreceptor outer segments, and secrete a wide range of growth factors and immunomodulatory molecules. Because of their critical roles in photoreceptor homeostasis and light-induced oxidative stress resistance, RPE1 cells are widely used as an in vitro model for studying retinal physiology, visual cycle metabolism, and retinal degenerative diseases such as age-related macular degeneration."
}

\vspace{0.15cm}

Based on the retrieved biological evidence, we perform prompt augmentation on the original query and construct a retrieval-augmented prompt as the input to the model. The enhanced prompt corresponding to the above example is shown below.

\vspace{0.15cm}
\noindent\textbf{Example of the augmented prompt.}
\vspace{0.15cm}

\noindent
{\itshape
"You are a molecular and cellular biology expert analyzing gene regulation upon CRISPR knockdown. Before you start reasoning, additional biological information will be provided to you, including functional summaries of the perturbed gene and the target gene, a description of the relevant cell line, and pathway-level mechanistic context linking the two genes. Use this additional context when forming your reasoning. First, provide your reasoning process within \texttt{<think>} \texttt{</think>} tags. Consider relevant pathways (e.g., cell-type specific biology, ribosome biogenesis, transcription, mitochondrial function, stress response), gene interactions, and cell-specific context. Then, choose one option from the following and place your choice within \texttt{<answer>}\texttt{</answer>} tags: 'upregulated', 'downregulated', or 'not significantly changed'. Example: \texttt{<think>} (Your reasoning here) \texttt{</think>} \texttt{<answer>} (upregulated / downregulated / not significantly changed) \texttt{<answer>}. Analyze the regulatory effect of knocking down the TULP1 gene on the ITCH gene in a single-cell RPE1 cell line using CRISPR interference. \\
\textbf{Perturbation\_Gene\_Information}: Perturbing TULP1 via gene knockdown disrupts photoreceptor synapse development and function, leading to impaired protein transport and phagocytosis in photoreceptor cells. This knockdown also affects the long-term survival of photoreceptor cells, potentially impacting eye function. Furthermore, the loss of TULP1 function may alter the organization of the cytoskeleton and lipid dynamics, given its interactions with these components. Additionally, the knockdown may influence the activity of genes involved in dendrite development, detection of light stimulus, and retina development and homeostasis, such as ARHGAP26, NRXN1, and TULP3. Overall, perturbing TULP1 may have far-reaching consequences for photoreceptor cell biology and visual perception. The primary molecular and cellular function of gene TULP1 is to regulate photoreceptor synapse development and function, as well as facilitate protein transport and phagocytosis in photoreceptor cells, through its interactions with cytoskeleton proteins and lipids. Additionally, TULP1 contributes to the long-term survival of photoreceptor cells. Perturbing gene TULP1 via gene knockdown may disrupt normal photoreceptor function and development, leading to impaired photoreceptor cell survival and potentially affecting eye function. \\
\textbf{Target\_Gene\_Information}: Perturbations that may impact the levels of ITCH include alterations in the activity of upstream regulators, such as SMAD3, or changes in the availability of its binding partners, including viral proteins and other cellular substrates. Additionally, perturbations in the formation or stability of protein complexes that ITCH participates in, such as the AIP4-DTX3L complex or the CPSF6-EWSR1-ITCH-NUDT21-POLR2A-UBAP2L complex, may also affect ITCH levels. Furthermore, changes in the ubiquitination or deubiquitination of ITCH itself, or its substrates, could also influence ITCH levels. The primary molecular and cellular function of gene ITCH is to act as an E3 ubiquitin-protein ligase, transferring ubiquitin to targeted substrates and facilitating their degradation, and to interact with various proteins and viruses to regulate their activity and fate. ITCH also participates in the formation of protein complexes that mediate ubiquitination and proteasomal degradation of specific targets. Perturbations that might impact the expression of gene ITCH include alterations in the activity of upstream regulators, such as SMAD3, or changes in the availability of its binding partners, including viral proteins and other cellular substrates. \\
\textbf{Cell\_Line\_Description}: The RPE1 cell line is derived from human retinal pigment epithelial tissue, a monolayer of hexagonal, melanin-containing cells that support photoreceptor function. These cells absorb stray light, maintain the blood–retina barrier, regulate ion and nutrient transport, recycle visual cycle components, phagocytose photoreceptor outer segments, and secrete a wide range of growth factors and immunomodulatory molecules. Because of their critical roles in photoreceptor homeostasis and light-induced oxidative stress resistance, RPE1 cells are widely used as an in vitro model for studying retinal physiology, visual cycle metabolism, and retinal degenerative diseases such as age-related macular degeneration. \\
\textbf{Pathway\_Context}: Pathway 1: TULP1 $\rightarrow$ photoreceptor inner segment $\rightarrow$ DRAM2 $\rightarrow$ apoptotic process $\rightarrow$ ITCH Pathway 2: TULP1 $\rightarrow$ phosphatidylinositol-4,5-bisphosphate binding $\rightarrow$ PFN1 $\rightarrow$ cell cortex $\rightarrow$ ITCH Pathway 3: TULP1 $\rightarrow$ photoreceptor inner segment $\rightarrow$ SAG $\rightarrow$ G protein-coupled receptor internalization $\rightarrow$ receptor internalization $\rightarrow$ ITCH \\
\textbf{Multi\_Modal\_Interaction\_Context}: To specifically analyze the relationship between the Perturbation Gene and the Target Gene, we provide three fused embedding vectors. These embeddings are generated via a Cross-Attention mechanism, capturing the interaction features between the two genes across different biological modalities. They are provided below, wrapped between Embedding\_Start and Embedding\_End tags: 1. Interactive Protein Sequence Features: Represents the physical structure compatibility and potential binding affinity changes. It highlights how the perturbation might physically impact the target protein structure. \texttt{<ESM-2\_EMB>} 2. Interactive Pathway Topology Context: Derived from the molecular pathway knowledge graph. It captures the structural signal transduction paths and upstream/downstream relationships between the genes. \texttt{<Path-KG\_EMB>} 3. Interactive Gene Regulatory Context: Derived from the gene-gene interaction network. It captures the regulatory distance, co-expression patterns, and network connectivity in the global regulatory landscape. \texttt{<Gene-KG\_EMB>} Please incorporate these specific interaction signals into your reasoning process."}

\vspace{0.15cm}

The above retrieval-augmented prompt is provided as input to AROMA, which generates the following reasoning output for this example.

\vspace{0.15cm}
\noindent\textbf{Example of the AROMA output.}
\vspace{0.15cm}

\noindent
{\itshape
"\texttt{<think>} TULP1 knockdown in RPE1 cells disrupts photoreceptor synapse development and function, leading to impaired protein transport and phagocytosis. This disruption may affect the organization of the cytoskeleton and lipid dynamics, potentially impacting the activity of genes involved in dendrite development, detection of light stimulus, and retina development and homeostasis. Although TULP1 is involved in regulating photoreceptor synapse development and function, its knockdown may not directly impact ITCH expression. ITCH's primary function is as an E3 ubiquitin-protein ligase, transferring ubiquitin to targeted substrates and facilitating their degradation, and interacting with various proteins and viruses to regulate their activity and fate. The known pathways connecting TULP1 to ITCH (Pathways 1, 2, and 3) do not provide a direct link between TULP1 knockdown and ITCH expression. Furthermore, the cellular context of RPE1 cells, which are derived from human retinal pigment epithelial tissue, may provide buffering or compensatory mechanisms that stabilize ITCH's expression, leading to a limited overall effect on its expression. Therefore, the knockdown of TULP1 in RPE1 cells is unlikely to significantly change ITCH expression.\texttt{</think>}\texttt{<answer>}not significantly changed\texttt{</answer>}."}

To further verify whether the model's reasoning process is genuinely grounded in the retrieved biological evidence, we conducted a sentence-level source-tracing analysis on the reasoning text generated by AROMA after obtaining its output. Specifically, this analysis aims to assess whether the reasoning content is supported by evidence that can be traced back to the retrieved knowledge sources. The results are presented as follows.

\vspace{0.15cm}
\noindent\textbf{Reasoning sentence 1.} \\
\textit{TULP1 knockdown in RPE1 cells disrupts photoreceptor synapse development and function, leading to impaired protein transport and phagocytosis.} \\
\textbf{Evidence source}: This is directly supported by \texttt{pert\_summary\_1} and \texttt{pert\_summary\_2}, which state that TULP1 knockdown disrupts photoreceptor synapse development and function and leads to impaired protein transport and phagocytosis in photoreceptor cells.

\vspace{0.15cm}
\noindent\textbf{Reasoning sentence 2.} \\
\textit{This disruption may affect the organization of the cytoskeleton and lipid dynamics, potentially impacting the activity of genes involved in dendrite development, detection of light stimulus, and retina development and homeostasis.} \\
\textbf{Evidence source}: This is grounded in \texttt{pert\_summary\_1}, which notes that loss of TULP1 function may alter cytoskeleton organization and lipid dynamics and may influence the activity of genes involved in dendrite development, light stimulus detection, and retina development and homeostasis.

\vspace{0.15cm}
\noindent\textbf{Reasoning sentence 3.} \\
\textit{Although TULP1 is involved in regulating photoreceptor synapse development and function, its knockdown may not directly impact ITCH expression.} \\
\textbf{Evidence source}: The first part is supported by \texttt{pert\_summary\_2}, which describes TULP1 as regulating photoreceptor synapse development and function. The judgment that TULP1 knockdown may not directly impact ITCH expression is an inference based on \texttt{gene\_summary\_1} and \texttt{gene\_summary\_2}, which list upstream regulators and binding partners of ITCH but do not mention TULP1 as a direct regulator.

\vspace{0.15cm}
\noindent\textbf{Reasoning sentence 4.} \\
\textit{ITCH's primary function is as an E3 ubiquitin-protein ligase, transferring ubiquitin to targeted substrates and facilitating their degradation, and interacting with various proteins and viruses to regulate their activity and fate.} \\
\textbf{Evidence source}: This sentence is directly supported by \texttt{gene\_summary\_2}, which describes ITCH as an E3 ubiquitin-protein ligase that transfers ubiquitin to substrates, facilitates their degradation, and interacts with various proteins and viruses to regulate their activity and fate.

\vspace{0.15cm}
\noindent\textbf{Reasoning sentence 5.} \\
\textit{The known pathways connecting TULP1 to ITCH (Pathways 1, 2, and 3) do not provide a direct link between TULP1 knockdown and ITCH expression.} \\
\textbf{Evidence source}: This is based on \texttt{pathway\_text}, which lists three multi-step paths from TULP1 to ITCH through intermediate compartments, binding events, and proteins; these paths indicate indirect, multi-hop connections rather than a direct regulatory link from TULP1 to ITCH expression.

\vspace{0.15cm}
\noindent\textbf{Reasoning sentence 6.} \\
\textit{Furthermore, the cellular context of RPE1 cells, which are derived from human retinal pigment epithelial tissue, may provide buffering or compensatory mechanisms that stabilize ITCH's expression, leading to a limited overall effect on its expression.} \\
\textbf{Evidence source}: This sentence is partially supported by the \texttt{cell\_line\_description}, which characterizes RPE1 as a retinal pigment epithelial cell line with multiple homeostasis-maintaining functions (such as phagocytosis of photoreceptor outer segments and secretion of growth factors and immunomodulatory molecules). The idea of buffering or compensatory mechanisms that stabilize ITCH expression is a higher-level inference based on this homeostatic context rather than an explicit statement in the retrieved text.

\vspace{0.15cm}
\noindent\textbf{Reasoning sentence 7.} \\
\textit{Therefore, the knockdown of TULP1 in RPE1 cells is unlikely to significantly change ITCH expression.} \\
\textbf{Evidence source}: This conclusion is an integrated inference from the retrieved evidence: TULP1 summaries focus on photoreceptor synapses, cytoskeleton, and lipid dynamics without indicating direct regulation of ITCH; ITCH summaries describe regulation by other upstream factors but do not mention TULP1; and the \texttt{pathway\_text} only provides indirect multi-hop connections from TULP1 to ITCH. Together with the homeostatic properties of RPE1 cells described in \texttt{cell\_line\_description}, these observations support the judgment that TULP1 knockdown is unlikely to strongly and directly alter ITCH expression.

\vspace{0.15cm}
\noindent\textbf{Final answer.} \\
\textit{\texttt{not significantly changed}} \\
\textbf{Evidence source}: This final categorical answer is consistent with the above reasoning, which suggests that, under the retrieved biological evidence and RPE1 cellular context, ITCH expression is not expected to undergo a strong or systematic change in response to TULP1 knockdown.

Overall, the sentence-level source-tracing results indicate that the key statements in the model-generated reasoning text are supported by corresponding biological evidence sources. This suggests that the reasoning produced by AROMA is not generated arbitrarily, but is instead constructed upon evidence-based reasoning paths, thereby enhancing the credibility and interpretability of the model's outputs.

\subsection{Biological Validity Analysis}

\begin{figure*}[t]
    \centering
    \includegraphics[width=\textwidth]{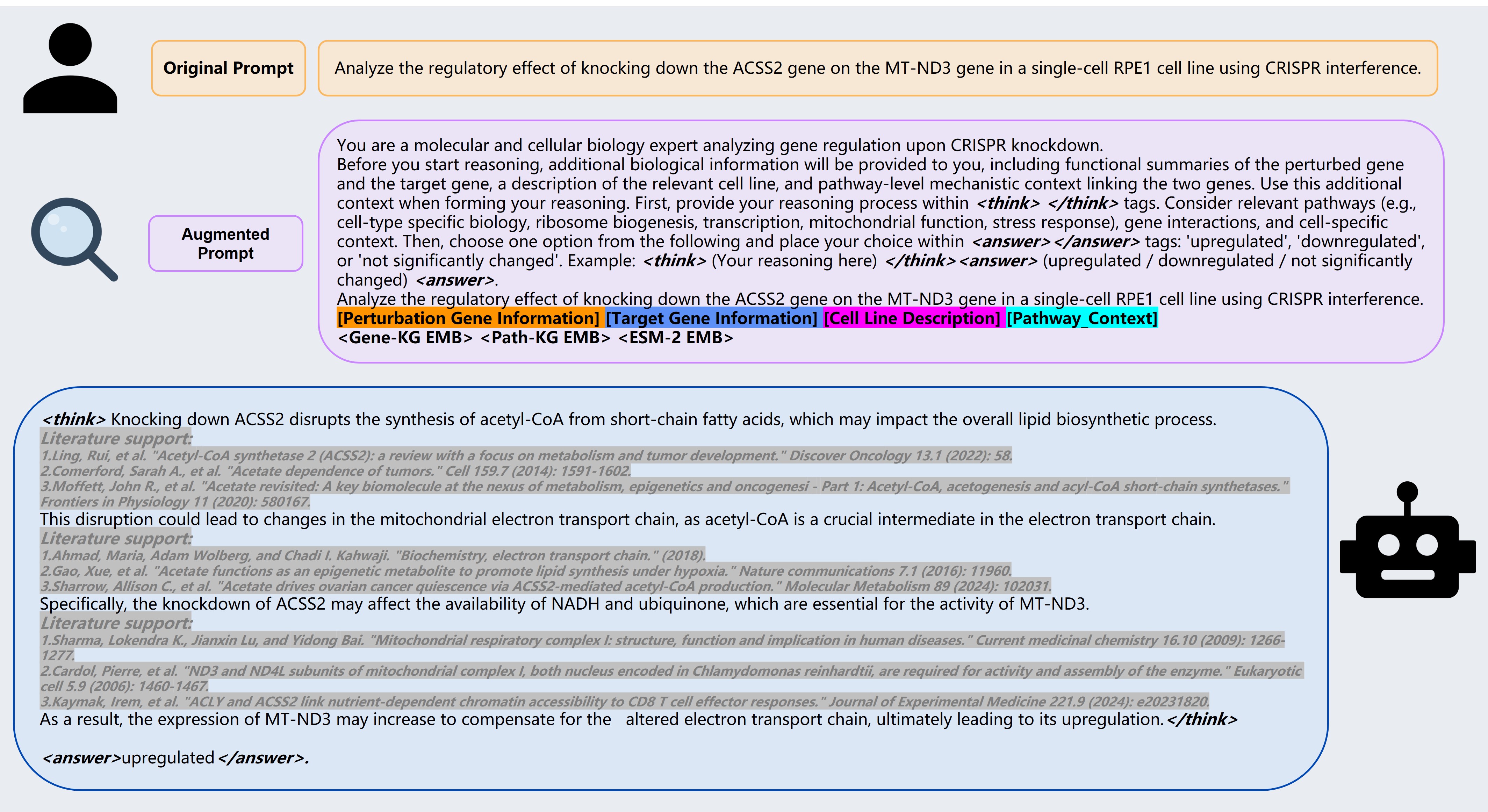}
    \caption{Visualization of the sentence-level biological validity analysis.}
    \label{fig:figure5}
\end{figure*}

To further verify the biological reliability and rationality of the model's generated content, this section analyzes the reasoning process produced by the model under a given genetic perturbation scenario, and examines whether its outputs are consistent with the experimental findings reported in relevant biological research papers. A visualization of the sentence-level biological validity analysis is provided in Figure~\ref{fig:figure5}.

Specifically, we consider the following case as the evaluation instance. 

\vspace{0.15cm}
\noindent\textbf{Example of Original Query.}
\vspace{0.15cm}

\noindent
{\itshape
"Analyze the regulatory effect of knocking down the ACSS2 gene on the MT-ND3 gene in a single-cell RPE1 cell line using CRISPR interference."
}
\vspace{0.15cm}

Based on this query, we retrieve biologically relevant evidence and construct an expanded retrieval-augmented prompt, which is then provided as input to the model. The resulting model-generated reasoning output is shown below.

\vspace{0.15cm}
\noindent\textbf{Example of the AROMA output.}
\vspace{0.15cm}

\noindent
{\itshape
"\texttt{<think>} Knocking down ACSS2 disrupts the synthesis of acetyl-CoA from short-chain fatty acids, which may impact the overall lipid biosynthetic process. This disruption could lead to changes in the mitochondrial electron transport chain, as acetyl-CoA is a crucial intermediate in the electron transport chain. Specifically, the knockdown of ACSS2 may affect the availability of NADH and ubiquinone, which are essential for the activity of MT-ND3. As a result, the expression of MT-ND3 may increase to compensate for the altered electron transport chain, ultimately leading to its upregulation.\texttt{</think>}\texttt{<answer>}upregulated\texttt{</answer>}."}

We choose this sample in this section for two main reasons. First, compared with cases labeled as “non-differential”, an upregulated label typically corresponds to more pronounced functional changes, making it more suitable for step-by-step mechanistic reasoning and causal pathway analysis. Second, this case simultaneously involves the linkage between metabolic pathways (ACSS2-mediated acetyl-CoA synthesis) and the mitochondrial electron transport chain (the complex containing MT-ND3), which requires the model to integrate multi-level biological knowledge during its reasoning process. Therefore, by carefully dissecting the reasoning process in this upregulated case, we can more directly assess whether the model’s inferred reasoning chain is aligned with the key biological mechanisms reported in experimental literature. The sentence-level analysis of the biological validity of this reasoning trace is presented below.

\noindent\textbf{Reasoning sentence 1.}\\
\textit{Knocking down ACSS2 disrupts the synthesis of acetyl-CoA from short-chain fatty acids, which may impact the overall lipid biosynthetic process.}

\vspace{0.15cm}
\noindent\textbf{Literature support:}
\vspace{0.1cm}

\noindent
\noindent\textbullet~A review on ACSS2 function \citep{ling2022acetyl} describes ACSS2 as acetyl-CoA synthetase short-chain family member 2, a nucleo-cytosolic enzyme that converts acetate and other short-chain fatty acids into acetyl-CoA for lipid metabolism and energy production.

\vspace{0.1cm}
\noindent
\noindent\textbullet~The same review further notes that ACSS2 is widely considered to activate a broad range of short-chain fatty acids into their corresponding acyl-CoA forms (e.g., butyrate, crotonate), although the catalytic efficiency can be cell-type dependent.

\vspace{0.1cm}
\noindent
\noindent\textbullet~A study in mammalian cells and mouse models \citep{comerford2014acetate} shows that knockdown or loss of ACSS2 decreases intracellular acetyl-CoA levels, reduces cytosolic protein acetylation, and induces autophagy, indicating that ACSS2 is a key source for maintaining the cellular acetyl-CoA pool.

\vspace{0.1cm}
\noindent
\noindent\textbullet~A study in hepatic disease models \citep{moffett2020acetate} reports that ACSS2 converts acetate into acetyl-CoA to fuel \textit{de novo} lipogenesis and triglyceride synthesis, and that ACSS2-deficient mice on a high-fat diet display markedly reduced hepatic lipid accumulation, consistent with impaired lipid biosynthetic capacity when ACSS2 is absent.

\vspace{0.1cm}
\noindent
\noindent\textbullet~A study in tumor cells \citep{comerford2014acetate} demonstrates that stable knockdown of ACSS2 significantly reduces incorporation of radiolabeled acetate into lipids, indicating that ACSS2-derived acetyl-CoA is essential for acetate-dependent lipid synthesis.

\vspace{0.1cm}
Taken together, these studies provide convergent evidence that ACSS2 plays a central role in generating short-chain-fatty-acid–derived acetyl-CoA, and that loss or knockdown of ACSS2 can disrupt acetate-dependent acetyl-CoA production and lipid biosynthetic processes.

\noindent\textbf{Reasoning sentence 2.}\\
\textit{This disruption could lead to changes in the mitochondrial electron transport chain, as acetyl-CoA is a crucial intermediate in the electron transport chain.}

\vspace{0.1cm}
\noindent\textbf{Literature support:}
\vspace{0.1cm}

\noindent\textbullet~A review on mitochondrial electron transport in StatPearls \citep{ahmad2018biochemistry} explains that the citric acid cycle produces NADH and FADH\textsubscript{2}, and that these reduced cofactors subsequently feed electrons into the electron transport chain, thereby linking TCA-cycle activity to respiratory chain function.

\vspace{0.1cm}
\noindent\textbullet~A metabolic tracing study in CD8\textsuperscript{+} T cells \citep{gao2016acetate} shows that acetate-derived acetyl-CoA contributes carbon both to cytosolic acetyl-CoA pools and to TCA cycle intermediates, and that acetate utilization can support mitochondrial oxidative phosphorylation and effector function.

\vspace{0.1cm}
\noindent\textbullet~A recent metabolic study in ovarian cancer cells \citep{sharrow2024acetate} reports that pharmacological inhibition of ACSS2 alters cellular metabolic programming, with reduced glycolytic activity and changes in mitochondrial oxygen consumption rates, suggesting that perturbation of ACSS2-dependent acetyl-CoA production can remodel mitochondrial respiratory chain function.

\vspace{0.1cm}
\noindent
Taken together, these studies support the notion that changes in ACSS2-dependent acetyl-CoA availability can indirectly influence the mitochondrial electron transport chain by modulating TCA-cycle-derived NADH and FADH\textsubscript{2}, the primary electron donors for respiratory electron transport.

\noindent\textbf{Reasoning sentence 3.}\\
\textit{Specifically, the knockdown of ACSS2 may affect the availability of NADH and ubiquinone, which are essential for the activity of MT-ND3.}
 
\vspace{0.1cm}
\noindent\textbf{Literature support:}
\vspace{0.1cm}

\noindent
\noindent\textbullet~A review on mitochondrial respiratory complex~I (NADH:ubiquinone oxidoreductase) \citep{sharma2009mitochondrial} summarizes that complex~I oxidizes matrix NADH and uses the released electrons to reduce ubiquinone to ubiquinol, thereby directly coupling NADH and ubiquinone availability to electron transfer activity in the respiratory chain. In parallel, MT-ND3 is described as a core subunit of complex~I that belongs to the minimal assembly required for catalyzing NADH dehydrogenation and electron transfer to ubiquinone \citep{cardol2006nd3}.

\vspace{0.1cm}
\noindent
\noindent\textbullet~Metabolic studies \citep{pryor2006free,kaymak2024acly} have shown that genetic knockdown or pharmacological inhibition of ACSS2 diminishes acetate-derived acetyl-CoA production and leads to reduced mitochondrial oxygen consumption and oxidative phosphorylation, indicating that ACSS2 activity supports efficient respiratory chain function.

\vspace{0.1cm}
Taken together, these findings support the notion that ACSS2-dependent metabolic activity can modulate mitochondrial electron transport by influencing NADH supply and ubiquinone-dependent complex~I function, providing a mechanistic basis for the proposed effect of ACSS2 knockdown on MT-ND3 activity.

Through literature-alignment analysis, we find that the reasoning chain generated by AROMA is consistently supported by relevant biological studies, with key reasoning statements receiving corroborating evidence from multiple levels, including metabolic pathways, mitochondrial function, and the electron transport chain. Overall, these results indicate that the model’s reasoning exhibits strong biological plausibility.

\section{KGs and Dataset Construction}
\label{app:data}

\subsection{Gene Knowledge Graph}

The Gene Knowledge Graph (Gene-KG) is used to characterize functional association structure at the gene level. We construct an undirected gene-gene association network by integrating the STRING and CORUM databases. Specifically, we first extract human protein-protein interactions from STRING and map protein identifiers to gene symbols. For any pair of genes with interaction evidence reported in STRING (e.g., experimental, curated database, or text-mining channels), we add an undirected edge between the two genes. We then collect human protein complex annotations from CORUM, expand each complex into pairwise connections among all of its constituent genes, and merge these edges into the same graph. To simplify the graph topology, we do not distinguish between different evidence sources or edge types; instead, all available evidence is collapsed into a single undirected association edge between the corresponding gene pair, and edges are interpreted as statistical associations rather than explicit causal regulatory relationships. After node normalization and duplicate-edge removal, the resulting Gene-KG contains 18{,}479 gene nodes and 752{,}612 edges, with an average degree of 81.46.

\subsection{Pathway Knowledge Graph}

The Pathway Knowledge Graph (Path-KG) is used to characterize functional associations and structural dependencies at the level of biological processes and pathways. We construct this graph by integrating resources from Gene Ontology and Reactome, and represent associations between genes and pathway or process nodes as an undirected graph. Concretely, from Gene Ontology (GO) we collect functional and process annotations for each gene and connect the gene to its associated GO term nodes. From Reactome, we further link each gene to the pathway modules or reaction units in which it participates, and merge the resulting pathway-level graph with the GO-based graph to obtain a unified pathway knowledge graph. To avoid unnecessary relation complexity, we do not distinguish between different relation sources or labels; instead, all functional participation and pathway membership information is collapsed into a single undirected association edge between the corresponding gene and pathway or ontology node. These edges are interpreted as structural and functional statistical associations rather than explicit causal regulatory dependencies. In addition, we perform denoising and structural regularization during preprocessing. On the one hand, we explicitly discard several overly generic high-level GO terms, such as GO:0005515 (protein binding) and GO:0003674 (molecular function), to reduce the impact of extremely broad concepts on graph statistics. On the other hand, in the merged graph we remove all nodes whose degree is greater than or equal to 1000, in order to suppress the dominance of a small number of hub nodes over local neighborhood structure and path distributions, and to obtain a topology that better reflects pathway-level semantic organization. After merging, deduplication, and hub filtering, the resulting Pathway-KG contains 80{,}020 nodes and 441{,}176 edges, with an average degree of 11.03. The basic statistics of the two constructed knowledge graphs are summarized in Table~\ref{tab:kg_stats}.

\subsection{PerturbReason Dataset}
In constructing the PerturbReason dataset, we first retrieve, for each perturbation--target--cell-line triplet, the relevant functional descriptions, path-based connectivity evidence, and cell-line background information from the knowledge graphs and external resources. These signals are organized into structured fields that provide contextual support for subsequent reasoning-trace generation. Below, we present a representative example of the raw retrieved information.

\begin{table}[H]
\centering
\small
\caption{Statistics of the constructed KGs.}
\label{tab:kg_stats}
\resizebox{\columnwidth}{!}{%
\begin{tabular}{lrrr}
\toprule
\textbf{Graph} & \textbf{Nodes} & \textbf{Edges} & \textbf{Avg. degree} \\
\midrule
Gene-KG    & 18{,}479 & 752{,}612 & 81.46 \\
Path-KG & 80{,}020 & 441{,}176 & 11.03 \\
\bottomrule
\end{tabular}%
}
\end{table}

\vspace{0.2cm}
\noindent\textbf{Example of retrieved information.}
\vspace{0.2cm}

\noindent
{\itshape
"\textbf{Perturbation\_gene}": "AAMP" \\
"\textbf{Target\_gene}": "AGPAT1" \\
"\textbf{Cell\_line}": "Jurkat" \\
"\textbf{Perturbation\_Gene\_summary\_1}": "Perturbing gene AAMP via gene knockdown may disrupt smooth muscle cell migration, potentially affecting the development and maintenance of vascular structures. This may also impact angiogenesis, as AAMP is involved in this process. Additionally, knockdown of AAMP may affect cell differentiation and endothelial cell migration, as it is involved in the positive regulation of these processes. Furthermore, AAMP's interactions with proteins such as EGFR, TBXA2R, and NOD2 may be disrupted, potentially influencing signaling pathways including EGFR dimerization, thromboxane signaling, and NOD1/2 signaling." \\
"\textbf{Perturbation\_Gene\_summary\_2}": "The primary molecular and cellular function of gene AAMP is to play a role in angiogenesis and cell migration, particularly in smooth muscle cell migration, where it may act through the RhoA pathway. AAMP's protein product, angio-associated migratory cell protein, interacts with other proteins such as AAMP and AEN to facilitate these processes.Perturbing gene AAMP via gene knockdown may disrupt smooth muscle cell migration, potentially affecting the development and maintenance of vascular structures." \\
"\textbf{Target\_Gene\_summary\_1}": "Perturbations that may impact the levels of AGPAT1 include alterations in the activity or expression of upstream regulators such as MEOX2 and PLCA, which could affect the transcriptional regulation of AGPAT1. Disruptions to the phosphatidic acid biosynthetic pathway, including changes in the availability of substrates such as lysophosphatidic acid, may also influence AGPAT1 expression through feedback mechanisms. Additionally, perturbations in the endoplasmic reticulum membrane, where AGPAT1 is localized, could affect the enzyme's activity and subsequent expression levels. Furthermore, changes in the expression or activity of physically interacting proteins, such as AGPAT2-5, LCLAT1, and MBOAT1-2, may also impact AGPAT1 levels." \\
"\textbf{Target\_gene\_summary\_2}": "The primary molecular and cellular function of gene AGPAT1 is to catalyze the conversion of lysophosphatidic acid (LPA) into phosphatidic acid (PA) through the incorporation of an acyl moiety at the sn-2 position of the glycerol backbone, thereby playing a crucial role in phosphatidic acid biosynthesis. This enzyme activity is essential for the regulation of lipid metabolism and membrane biogenesis.Perturbations that might impact the expression of gene AGPAT1 include alterations in the levels or activity of upstream regulators, such as MEOX2 and PLCA, or disruptions to the phosphatidic acid biosynthetic pathway, which could affect the availability of substrates or the feedback regulation of AGPAT1 expression." \\
"\textbf{Pathway\_text}": "Pathway 1: AAMP $\rightarrow$ cytosol $\rightarrow$ ABHD5 $\rightarrow$ 1-acylglycerol-3-phosphate O-acyltransferase activity $\rightarrow$ AGPAT1. Pathway 2: AAMP $\rightarrow$ cytosol $\rightarrow$ ABHD5 $\rightarrow$ lysophosphatidic acid acyltransferase activity $\rightarrow$ 1-acylglycerol-3-phosphate O-acyltransferase activity $\rightarrow$ AGPAT1. Pathway 3: AAMP $\rightarrow$ positive regulation of endothelial cell migration $\rightarrow$ ZNF580 $\rightarrow$ positive regulation of interleukin-8 production $\rightarrow$ positive regulation of cytokine production $\rightarrow$ AGPAT1." \\
"\textbf{Cell\_line\_description}": "Jurkat is a human T lymphocyte cell line derived from the peripheral blood of a patient with acute T-cell leukemia. It is widely used as an in vitro model to study T-cell receptor (TCR) signaling, activation, interleukin-2 (IL-2) production, and viral infection mechanisms such as HIV. Jurkat cells grow in suspension, are easily transfected, and can produce IL-2 upon TCR stimulation. They serve as an essential tool for investigating intracellular pathways involving calcium signaling, protein kinase C, and NF-$\kappa$B activation. Due to specific mutations in genes such as PTEN and INPP5D, Jurkat cells show altered signal transduction and genomic stability, which should be considered when interpreting experimental results."}

\vspace{0.2cm}

After obtaining the above structured evidence, we convert it into an instruction-style contextual input and explicitly constrain the model to generate outputs solely based on the provided information. At the same time, the model is required to produce a reasoning chain that is consistent with the ground-truth label and remains well-formatted and semantically coherent. Below we provide a prompt example constructed from the above illustrative sample.

\vspace{0.2cm}
\noindent\textbf{Example of Prompt for Generating Reasoning Data.}
\vspace{0.2cm}

\noindent
{\itshape
"You are creating a reasoning dataset for molecular biology. The task is to generate a causal reasoning chain explaining why the knockdown of AAMP in jurkat leads to the observed expression pattern of AGPAT1. All reasoning must be based solely on the given knowledge below. Even when evidence is incomplete, you must construct a self-consistent reasoning chain that logically supports the provided label. The final output must strictly follow the dataset format and end exactly with: \texttt{<answer>}not significantly changed\texttt{</answer>}.\\
\textbf{Known Information}:\\
\textbf{Cell line}: Jurkat\\
\textbf{Cell line description}: Jurkat is a human T lymphocyte cell line derived from the peripheral blood of a patient with acute T-cell leukemia. It is widely used as an in vitro model to study T-cell receptor (TCR) signaling, activation, interleukin-2 (IL-2) production, and viral infection mechanisms such as HIV. Jurkat cells grow in suspension, are easily transfected, and can produce IL-2 upon TCR stimulation. They serve as an essential tool for investigating intracellular pathways involving calcium signaling, protein kinase C, and NF-$\kappa$B activation. Due to specific mutations in genes such as PTEN and INPP5D, Jurkat cells show altered signal transduction and genomic stability, which should be considered when interpreting experimental results.\\
\textbf{Perturbation gene}: AAMP\\
\textbf{Target gene}: AGPAT1\\
\textbf{Functional summary of AAMP}: Perturbation gene AAMP via gene knockdown may disrupt smooth muscle cell migration, potentially affecting the development and maintenance of vascular structures. This may also impact angiogenesis, as AAMP is involved in this process. Additionally, knockdown of AAMP may affect cell differentiation and endothelial cell migration, as it is involved in the positive regulation of these processes. Furthermore, AAMP's interactions with proteins such as EGFR, TBXA2R, and NOD2 may be disrupted, potentially influencing signaling pathways including EGFR dimerization, thromboxane signaling, and NOD1/2 signaling. The primary molecular and cellular function of gene AAMP is to play a role in angiogenesis and cell migration, particularly in smooth muscle cell migration, where it may act through the RhoA pathway. AAMP's protein product, angio-associated migratory cell protein, interacts with other proteins such as AAMP and AEN to facilitate these processes. Perturbation gene AAMP via gene knockdown may disrupt smooth muscle cell migration, potentially affecting the development and maintenance of vascular structures.\\
\textbf{Functional summary of AGPAT1}: Perturbations that may impact the levels of AGPAT1 include alterations in the activity or expression of upstream regulators such as MEOX2 and PLCA, which could affect the transcriptional regulation of AGPAT1. Disruptions to the phosphatidic acid biosynthetic pathway, including changes in the availability of substrates such as lysophosphatidic acid, may also influence AGPAT1 expression through feedback mechanisms. Additionally, perturbations in the endoplasmic reticulum membrane, where AGPAT1 is localized, could affect the enzyme's activity and subsequent expression levels. Furthermore, changes in the expression or activity of physically interacting proteins, such as AGPAT2-5, LCLAT1, and MBOAT1-2, may also impact AGPAT1 levels. The primary molecular and cellular function of gene AGPAT1 is to catalyze the conversion of lysophosphatidic acid (LPA) into phosphatidic acid (PA) through the incorporation of an acyl moiety at the sn-2 position of the glycerol backbone, thereby playing a crucial role in phosphatidic acid biosynthesis. This enzyme activity is essential for the regulation of lipid metabolism and membrane biogenesis. Perturbations that might impact the expression of gene AGPAT1 include alterations in the levels or activity of upstream regulators, such as MEOX2 and PLCA, or disruptions to the phosphatidic acid biosynthetic pathway, which could affect the availability of substrates or the feedback regulation of AGPAT1 expression.\\
\textbf{Known pathway}: Pathway 1: AAMP → cytosol → ABHD5 → 1-acylglycerol-3-phosphate O-acyltransferase activity → AGPAT1. Pathway 2: AAMP → cytosol → ABHD5 → lysophosphatidic acid acyltransferase activity → 1-acylglycerol-3-phosphate O-acyltransferase activity → AGPAT1. Pathway 3: AAMP → positive regulation of endothelial cell migration → ZNF580 → positive regulation of interleukin-8 production → positive regulation of cytokine production → AGPAT1. \\
\textbf{Output Requirements}: 1.Use only the 'Known Information'. No external knowledge or assumptions. 2.The reasoning \texttt{<think>} must form a full causal chain that is logically consistent with the final label. 3.If intermediate evidence is missing, explain the outcome through plausible buffering, compensation, or feedback effects. 4.The conclusion (\texttt{<answer>}) must exactly match 'not significantly changed'. 5.Output format must be: \texttt{<think>}...\texttt{</think>}\texttt{<answer>}not significantly changed\texttt{</answer>}. Example (for format reference only, not to be copied): \texttt{<think>}Knocking down AAMP may weaken the pathway regulating several downstream transcription factors. Although some mediators are missing, compensatory mechanisms in jurkat could stabilize AGPAT1's expression, leading to a limited overall effect.\texttt{</think>}\texttt{<answer>}not significantly changed\texttt{</answer>} Reminder: The model output will be used directly as reasoning dataset entries. Therefore, ensure your final answer strictly ends with \texttt{<answer>}not significantly changed\texttt{</answer>} and follows the dataset format."}

Given the above prompt, we query the DeepSeek-R1-0528 model to generate the reasoning data. The \texttt{<think>} segment represents the reasoning process formed by the model under the provided evidence constraints, while the \texttt{<answer>} segment corresponds to the final three-way classification output. An example of the model-generated result for this sample is shown below:

\vspace{0.2cm}
\noindent\textbf{Example of Generated Reasoning Data.}
\vspace{0.2cm}

\noindent
{\itshape
"\texttt{<think>} AAMP's primary function is to regulate smooth muscle cell migration, angiogenesis, and cell differentiation. In Jurkat cells, AAMP's knockdown may disrupt these processes, potentially affecting the development and maintenance of vascular structures. However, AGPAT1 is primarily involved in phosphatidic acid biosynthesis, which is not directly related to AAMP's functions. Although AAMP is involved in the positive regulation of endothelial cell migration, this pathway is not directly linked to AGPAT1's expression. The known pathways connecting AAMP to AGPAT1 (Pathways 1-3) involve intermediate enzymes and proteins, which may buffer the effects of AAMP knockdown. In Jurkat cells, the absence of AAMP may lead to compensatory mechanisms that stabilize AGPAT1's expression, resulting in a limited overall effect on its expression levels. Furthermore, the cell line's specific mutations in genes like PTEN and INPP5D may also influence the outcome, potentially mitigating the impact of AAMP knockdown on AGPAT1. Therefore, it is likely that the knockdown of AAMP in Jurkat cells will not significantly change the expression of AGPAT1. \texttt{</think>}." \texttt{<answer>}not significantly changed\texttt{</answer>}."}

\section{Quality Analysis of the Dataset}
\label{app:reasoning_quality}

To more systematically evaluate the overall quality of the reasoning data and to further verify that the generated reasoning content is indeed grounded in biological evidence retrieved from our knowledge base, we conduct a source-tracing consistency analysis on the reasoning text of the above example. Specifically, we decompose the model-generated reasoning output into several key statements, and examine, on a sentence-level basis, whether each of these statements is supported by the retrieved evidence in terms of content, logical direction, and biological interpretation, aligning each statement with its corresponding evidence source. In this way, we are able not only to assess whether individual reasoning statements are evidence-supported, but also to observe whether the model makes consistent and coherent use of evidence along the entire reasoning chain. The detailed sentence-level alignment and evidence-matching results are presented below.

\medskip
\noindent\textbf{Reasoning sentence 1.} 
\emph{``AAMP's primary function is to regulate smooth muscle cell migration, angiogenesis, and cell differentiation.''}\\
\textbf{Evidence source:} Supported by \texttt{pert\_summary\_1} and \texttt{pert\_summary\_2}, which describe AAMP as participating in smooth muscle cell migration, angiogenesis, and cell differentiation.

\noindent\textbf{Reasoning sentence 2.} 
\emph{``In Jurkat cells, AAMP's knockdown may disrupt these processes, potentially affecting the development and maintenance of vascular structures.''}\\
\textbf{Evidence source:} Supported by \texttt{pert\_summary\_1} , which states that AAMP knockdown may disrupt smooth muscle cell migration and affect vascular structure development and maintenance, in the context of the Jurkat cell line.

\noindent\textbf{Reasoning sentence 3.} 
\emph{``However, AGPAT1 is primarily involved in phosphatidic acid biosynthesis, which is not directly related to AAMP's functions.''}\\
\textbf{Evidence source:} Grounded in \texttt{gene\_summary\_2}, which characterizes AGPAT1 as playing a key role in phosphatidic acid biosynthesis, in contrast to the migration- and angiogenesis-related functions of AAMP described in \texttt{pert\_summary\_1} and \texttt{pert\_summary\_2}.

\noindent\textbf{Reasoning sentence 4.} 
\emph{``Although AAMP is involved in the positive regulation of endothelial cell migration, this pathway is not directly linked to AGPAT1’s expression.''}\\
\textbf{Evidence source:} Supported by \texttt{pathway\_text} (Pathway~3), where AAMP connects to AGPAT1 through multiple intermediate nodes such as ZNF580 and cytokine-related regulatory steps, indicating an indirect multi-hop pathway rather than a direct expression link.

\noindent\textbf{Reasoning sentence 5.} 
\emph{``The known pathways connecting AAMP to AGPAT1 (Pathways 1--3) involve intermediate enzymes and proteins, which may buffer the effects of AAMP knockdown.''}\\
\textbf{Evidence source:} Based on \texttt{pathway\_text} (Pathways~1--3), which contain intermediate enzymes and regulatory proteins including ABHD5 and acyltransferase-related activities, showing that the linkage between AAMP and AGPAT1 passes through multiple intermediary steps.

\noindent\textbf{Reasoning sentence 6.} 
\emph{``In Jurkat cells, the absence of AAMP may lead to compensatory mechanisms that stabilize AGPAT1's expression, resulting in a limited overall effect on its expression levels.''}\\
\textbf{Evidence source:} Informed by \texttt{gene\_summary\_1} and \texttt{gene\_summary\_2}, which note upstream regulatory and feedback mechanisms affecting AGPAT1 expression, together with \texttt{cell\_line\_description}, which reports altered signaling in Jurkat cells.

\noindent\textbf{Reasoning sentence 7.} 
\emph{``Furthermore, the cell line's specific mutations in genes like PTEN and INPP5D may also influence the outcome, potentially mitigating the impact of AAMP knockdown on AGPAT1.''}\\
\textbf{Evidence source:} Directly supported by \texttt{cell\_line\_description}, which states that Jurkat cells harbor mutations in PTEN and INPP5D and exhibit altered signal transduction.

\noindent\textbf{Reasoning sentence 8.} 
\emph{``Therefore, it is likely that the knockdown of AAMP in Jurkat cells will not significantly change the expression of AGPAT1.''} (answer: \emph{not significantly changed})\\
\textbf{Evidence source:} Synthesized from the provided functional, pathway, and cellular evidence.

Overall, this source-tracing analysis confirms that the generated reasoning is directly supported by the retrieved evidence, ensuring the reliability and quality of the constructed dataset.

\end{document}